\documentclass[english,aps, twocolumn, prresearch, longbibliography, superscriptaddress]{revtex4-2}
\usepackage[T1]{fontenc}
\usepackage[latin9]{inputenc}
\usepackage{color}
\usepackage{babel}
\usepackage{amsmath}
\usepackage{graphicx}
\usepackage[unicode=true,pdfusetitle,
 bookmarks=true,bookmarksnumbered=false,bookmarksopen=false,
 breaklinks=false,pdfborder={0 0 0},pdfborderstyle={},backref=false,colorlinks=true]
 {hyperref}
\hypersetup{
 linkcolor=darkblue, citecolor=darkblue, urlcolor=darkblue}

\makeatletter

\usepackage{color}
\usepackage{babel}
\usepackage{breakurl}
\usepackage{orcidlink}

\renewcommand{\fnum@figure}{FIG.~\thefigure}

\definecolor{darkblue}{rgb}{0.0, 0.0, 0.55}

\@ifundefined{showcaptionsetup}{}{
 \PassOptionsToPackage{caption=false}{subfig}}

\usepackage[caption=false]{subfig} 

\makeatother

\begin{document}
\title{Autoresonant excitation of space-time quasicrystals in plasma}
\author{Vadim R. Munirov\orcidlink{0000-0001-6711-1272}}
\email{vmunirov@berkeley.edu}

\affiliation{Department of Physics, University of California, Berkeley, California
94720, USA}
\author{Lazar Friedland\orcidlink{0000-0002-3603-6908}}
\affiliation{Hebrew University of Jerusalem, Jerusalem 91904, Israel}
\author{Jonathan S. Wurtele\orcidlink{0000-0001-8401-0297}}
\affiliation{Department of Physics, University of California, Berkeley, California
94720, USA}
\date{January 6, 2022; 
PRR: received 7 January 2022; accepted 19 March 2022; published May 25 2022}
\begin{abstract}
We demonstrate theoretically and numerically that a warm fluid model
of a plasma supports space-time quasicrystalline structures. These
structures are highly nonlinear, two-phase, ion acoustic waves that
are excited autoresonantly when the plasma is driven by two small
amplitude chirped-frequency ponderomotive drives. The waves exhibit
density excursions that substantially exceed the equilibrium plasma
density. Remarkably, these extremely nonlinear waves persist even
when the small amplitude drives are turned off. We derive the weakly
nonlinear analytical theory by applying Whitham\textquoteright s averaged
variational principle to the Lagrangian formulation of the fluid equations.
The resulting system of coupled weakly nonlinear equations is shown
to be in good agreement with fully nonlinear simulations of the warm
fluid model. The analytical conditions and thresholds required for
autoresonant excitation to occur are derived and compared to simulations.
The weakly nonlinear theory guides and informs numerical study of
how the two-phase quasicrystalline structure ``melts'' into a single
phase traveling wave when one drive is below a threshold. These nonlinear
structures may have applications to plasma photonics for extremely
intense laser pulses, which are limited by the smallness of density
perturbations of linear waves.
\end{abstract}
\maketitle

\section{Introduction}

Photonic crystals \citep{Joannopoulos2008,Ho1990,John1987,Yablonovitch1987}
built from conventional materials are routinely employed to focus,
polarize, and manipulate light pulses. A periodic array of alternating
dielectrics and plasmas called plasma photonic crystals (PPCs) has
been the subject of much interest \citep{Hojo2004,Hojo2009,Solaimani2020,Limei2010,Rahmani2019,Li2009,Zhang2014,Sakai2005,Sakai2012,Qi2015,Zhang2013,Zhang2013b,Suzuki2014,Liang2021,Wang2016}
owing to its optical properties. Besides crystals, Levine and Steinhardt
\citep{Levine1984} introduced quasicrystals \textemdash{} materials
with properties that are ordered in space but do not possess an exact
periodicity. Photonic quasicrystals have been studied extensively
\citep{Dyachenko2012,Poddubny2010,Lifshitz2005,Vardeny2013,Bahabad2008,Freedman2006,NeveOz2010,Chan1998}.
Moreover, crystals can be periodic not just in space but also in time.
For example, the optical properties of photonic time crystals with
a refractive index varying periodically in time have been investigated
in Refs.~\citep{Lustig2018,ZuritaSanchez2009,ZuritaSanchez2012,ZuritaSanchez2010,Salem2015,MartinezRomero2018,Sacha2020,Wang2018}.
The optical properties of the structures possessing periodicity both
in space and time have been studied in Refs.~\citep{Wang2018,Park2021,Biancalana2007,Biancalana2008,Zeng2017,Caloz2020,Caloz2020b,Sharabi2020}.

Plasma photonic crystals have a major drawback where the plasmas are
contained within solid material: they break down at high field intensity
and are, therefore, incapable of controlling the laser pulses that
are essential for many high energy density science applications. Purely
plasma based structures, on the other hand, can withstand high intensity
pulses. Many concepts for plasma-based optical elements have been
proposed and built \citep{Milchberg2019}. Plasma channels \citep{DurfeeMilchberg1993}
routinely focus lasers for particle acceleration experiments. Over
two decades ago, it was realized that resonant interactions \citep{Malkin1999}
could be used for compression of intense pulses (replacing large compressor
gratings). Plasma mirrors \citep{Thaury2007,Doumy2004,Dromey2004,Kapteyn1991}
are routinely used \citep{Michel2009,Michel2009b} at the National
Ignition Facility (NIF) to improve performance in inertial fusion
experiments. Laser-sculpted plasma grating structures \citep{Lehmann2016,Lehmann2017}
as well as polarization control using plasma structures \citep{Lehmann2018,Michel2014,Michel2020,Kur2021}
have been investigated. A significant challenge of plasma gratings
is that their efficacy depends on the maximum variation in the index
of refraction that can be achieved at the required spatial scale.
Plasma density is well below critical density, and the density modulation
should be as large as possible. This naturally leads to an exploration
of nonlinear waves.

In this paper, we propose formation and control of space-time quasicrystalline
structures in plasma through excitation of strongly nonlinear large
amplitude multiphase ion acoustic waves that modulate plasma density
in the desired way. It is known that a linear standing wave of the
form $U(x,t)\propto\cos(kx)\cos(\omega t)$, which is periodic in
both time and space, is formed by superposing two linear traveling
waves of the same frequency $\omega$, but propagating in the opposite
direction. Nonlinearity of the media in some cases allows a generalization
of linear standing waves to a waveform $U(x,t)=F\left(kx-\omega t,kx+\omega t\right)$,
where $F$ is a $2\pi$-periodic nonlinear function of two phase variables
and is also periodic in time and space. Recently, it was demonstrated
that such nonlinear structures could be formed in plasma in the form
of electron plasma \citep{Friedland2020} and ion acoustic \citep{Friedland2019}
waves. It is also known that even more complex multiphase constructs
of the form $U(x,t)=F\left(\theta_{1},\theta_{2},\ldots,\theta_{N}\right),$
where each phase is $\theta_{i}=k_{i}x-\omega_{i}t$, exist in other
physical systems described by integrable nonlinear wave equations,
such as the Korteweg-de Vries (KdV), nonlinear Schr\"{o}dinger (NLS),
and sine-Gordon (SG) equations \citep{Scott1999}. Generally, multiphase
functions $F$ are very nontrivial and can be described by a complex
analysis based on the Inverse Scattering Transform (see, for example,
Ref.~\citep{Novikov1984} for the KdV case). The multiphase waves
are $2\pi$-periodic in each of the phase variables, but if at least
two of $k_{i}$ or $\omega_{i}$ are not commensurate, function $F$
is not exactly periodic in time and/or space and, thus, comprise a
family of nontrivial space-time quasicrystals. But how can one excite
such a multiphase wave in a given physical system? Because of the
complexity of the waveform, a direct realization of a multiphase wave
requires setting up precise initial conditions, making it an impractical,
if not impossible, strategy for experiments even in the case of just
two phases $\theta_{1}$, $\theta_{2}$. A possible way to circumvent
the experimental difficulty of setting up the precise initial conditions
is exploiting phase locking with small amplitude chirped-frequency
traveling waves. In the past this approach was used in exciting multiphase
solutions for integrable systems: KdV \citep{Friedland2003}, NLS
\citep{Friedland2005}, SG \citep{Shagalov2009}, and the periodic
Toda lattice \citep{Khasin2003}. The autoresonant excitation was
also demonstrated for single phase large amplitude ion acoustic waves
\citep{Friedland2014,Friedland2017}. This suggests that perhaps the
autoresonance could be used in a more general way to excite multiphase
solutions for ion acoustic waves.

\noindent 
\begin{figure*}[tp]
\textbf{(a)}

\includegraphics[width=1\textwidth]{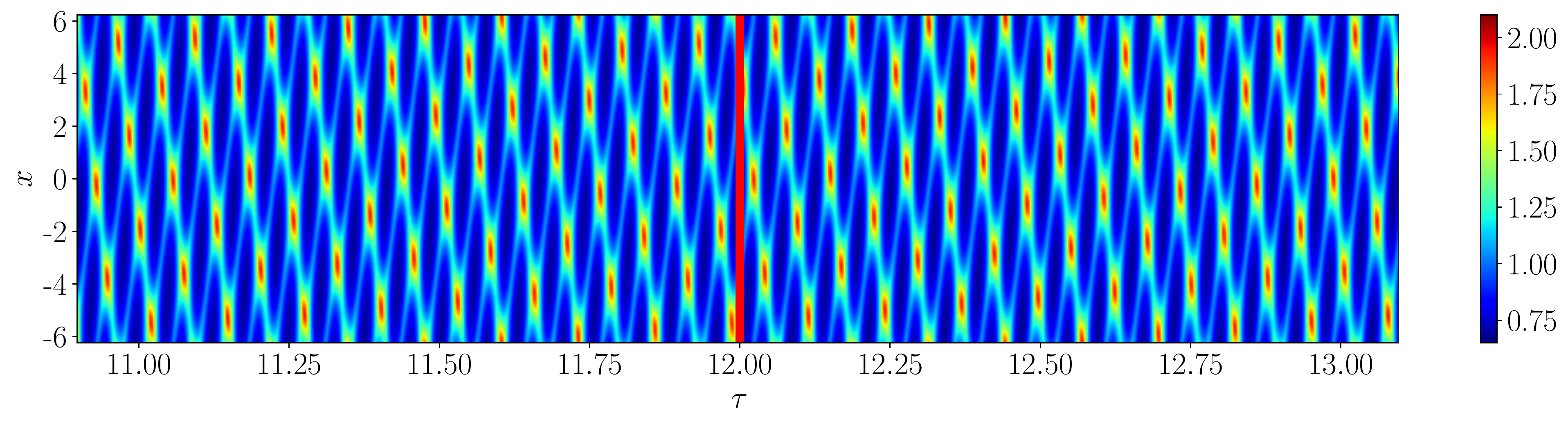}

\textbf{(b)}

\includegraphics[width=1\textwidth]{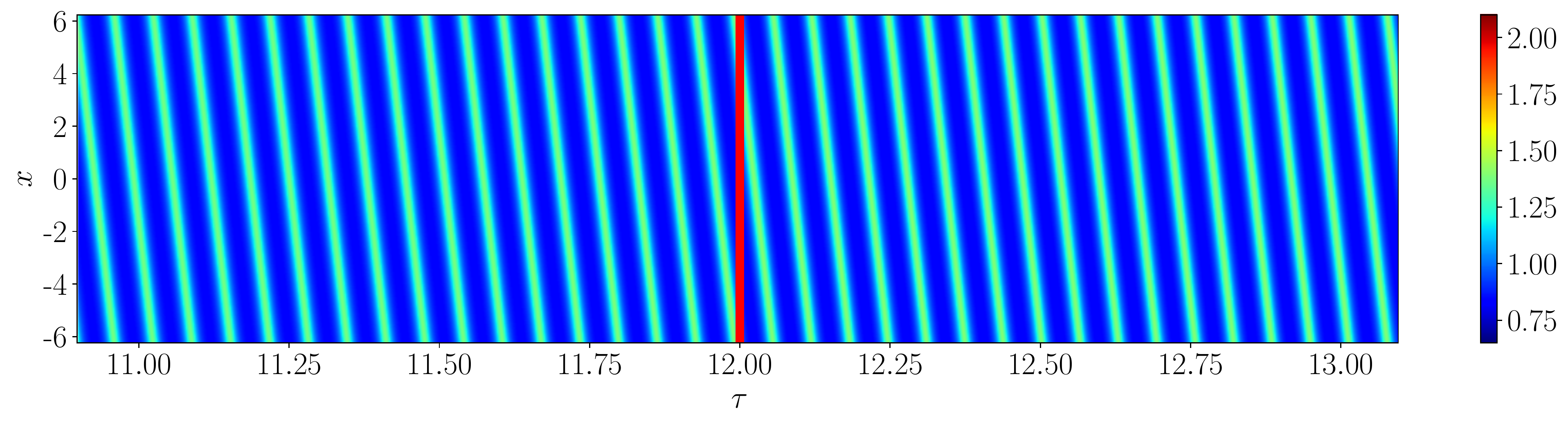}

\textbf{(c)}

\includegraphics[width=1\textwidth]{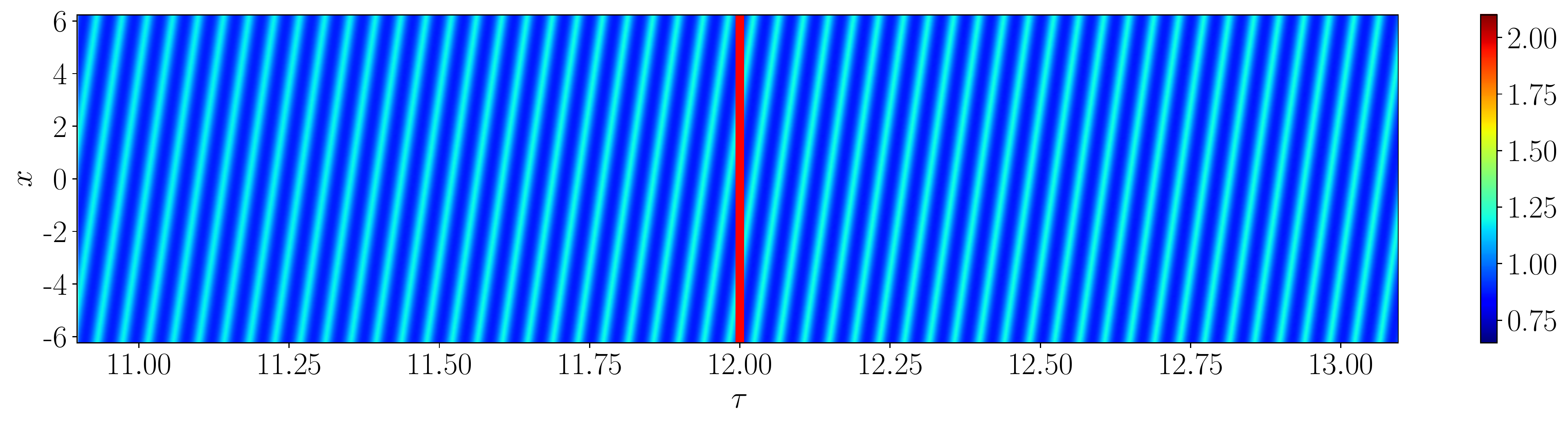}

\caption{\label{fig_full}The colormap of the electron density $n_{e}\left(x,\tau\right)$
as a function of slow time $\tau=\sqrt{\alpha_{1}}t$ and coordinate
$x$. (a) Two-phase autoresonant ion acoustic wave excited by two
driving counter propagating traveling waves with $k_{1}=-0.5$ and
$k_{2}=1$ obtained by solving the fully nonlinear equations~(\ref{eq:n_t})\textendash (\ref{eq:phi_xx}).
(b) Autoresonant single phase ion acoustic wave driven by the first
driving component in (a) with $k_{1}=-0.5$. (c) Autoresonant single
phase ion acoustic wave driven by the second driving component in
(a) with $k_{2}=1$. The red vertical line indicates the termination
of the drive at $\tau=12$.}
\end{figure*}

Here, we demonstrate, using a warm fluid plasma model, that a two-phase,
strongly nonlinear ion acoustic wave can be generated and controlled.
The wave is created by starting from zero and autoresonantly driving
the system with two small amplitude chirped-frequency ponderomotive
traveling waves. We show that slow passage of the driving waves through
resonances in the plasma results in the continuing autoresonant (phase
locked with both drives) excitation of the wave. The system sustains
this double autoresonance as the driving frequencies vary in time
by increasing the amplitude of the excited waveform, creating an extremely
large amplitude space-time quasicrystal in plasma. This result is
surprising and suggests some degree of integrability in the problem.
We conjecture that this integrability is related to the fact that
in the limit of small amplitude, ion acoustic waves are reduced to
the KdV-type equation \citep{Washimi1966}. Since there exist many
continuous physical systems approximated by the KdV, NLS, and SG equations,
we expect that autoresonant space-time quasicrystals can be formed
similarly in all these systems.

The paper is organized as follows. In Sec.~\ref{sec_II}, we formulate
the problem within a warm fluid plasma model and present a fully nonlinear
numerical solution. In Sec.~\ref{sec_III}, by using the Lagrangian
formulation of the fluid equations and applying Whitham\textquoteright s
averaged variational principle \citep{Whitham1999}, we derive an
analytical weakly nonlinear theory and demonstrate that it agrees
well with the fully nonlinear simulations. In Sec.~\ref{sec:IV},
we use our weakly nonlinear theory to study how to choose the parameters
required to excite a two-phase autoresonant solution as well as discuss
its threshold nature. Finally, in Sec.~\ref{sec_Conclusion}, we
summarize and discuss our results.

\noindent 
\begin{figure}
\includegraphics[width=1\columnwidth]{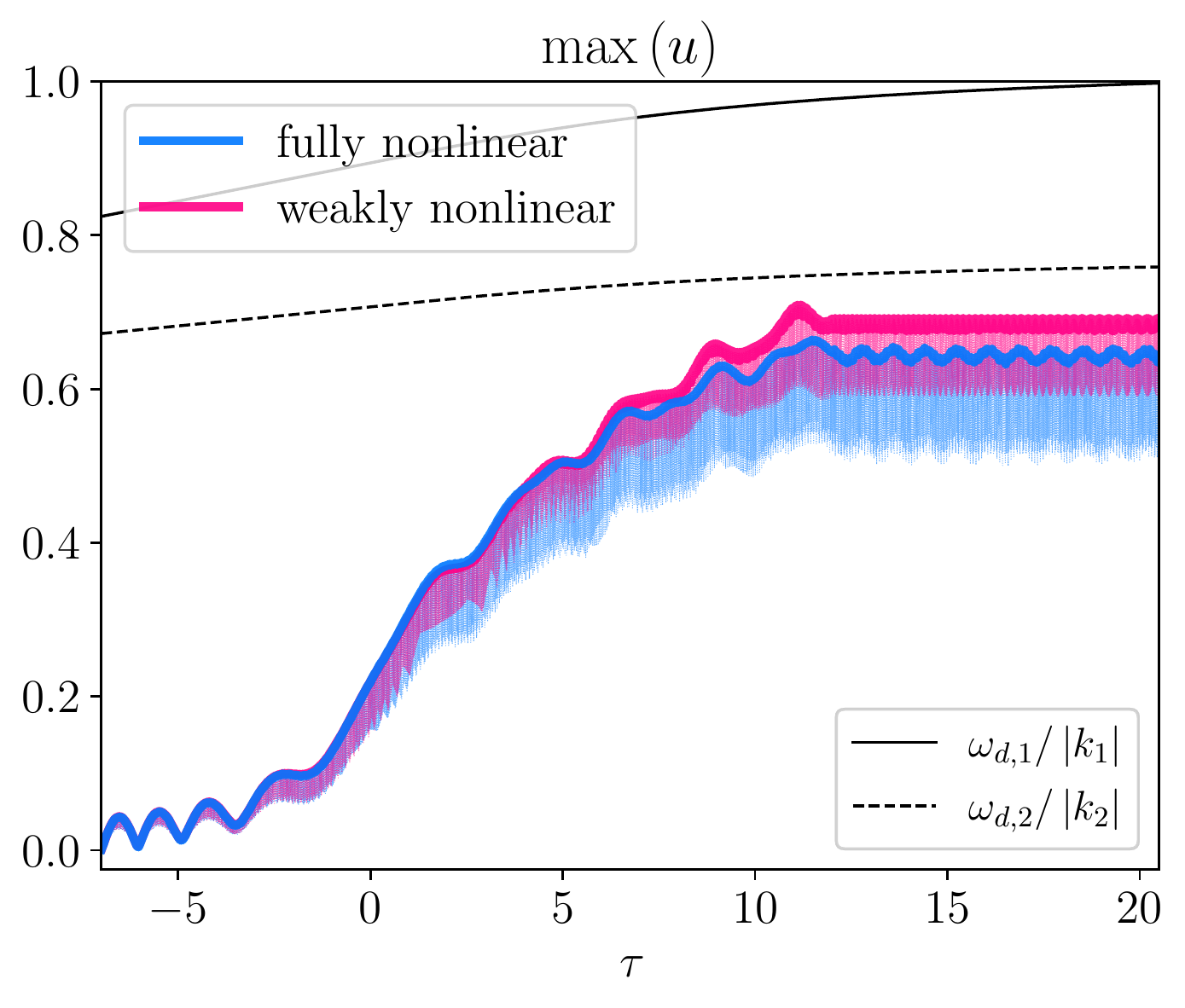}

\caption{\label{fig_max_u}The maximum (over $x$) of the ion fluid velocity
$u\left(x,\tau\right)$ versus slow time $\tau=\sqrt{\alpha_{1}}t$
for a two-phase autoresonant ion acoustic wave excited by two driving
traveling waves with $k_{1}=-0.5$ and $k_{2}=1$ obtained by solving
the fully nonlinear equations~(\ref{eq:n_t})\textendash (\ref{eq:phi_xx})
(denoted as \textquotedblleft fully nonlinear\textquotedblright ,
blue line) and the weakly nonlinear reduced dynamical equations~(\ref{eq:dI1_dt})\textendash (\ref{eq:dPHI2_dt})
(denoted as \textquotedblleft weakly nonlinear\textquotedblright ,
pink line). The solid black line represents the absolute value of
the phase velocity $\omega_{d,1}\left(\tau\right)/\left|k_{1}\right|$
of the first driving wave and the dashed black line represents the
absolute value of the phase velocity of the second driving wave $\omega_{d,2}\left(\tau\right)/\left|k_{2}\right|$
versus $\tau$. The parameters used in the simulations are the same
as in Figs.~\hyperref[fig_full]{\ref{fig_full}(a)},~\ref{fig_weakly},
and~\ref{fig_I_Phi_eps=00003D1.0}.}
\end{figure}

\section{\label{sec_II}Formulation of the problem and the numerical results}

\begin{figure*}[t]
\includegraphics[width=1\textwidth]{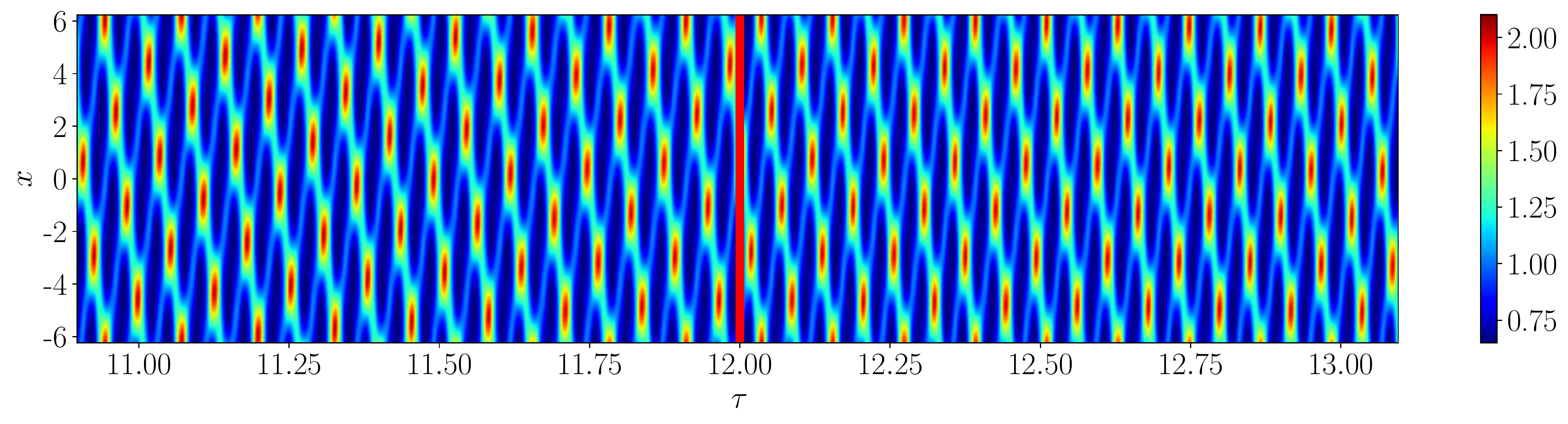}

\caption{\label{fig_weakly}The colormap of the electron density $n_{e}\left(x,\tau\right)$
as a function of slow time $\tau=\sqrt{\alpha_{1}}t$ and coordinate
$x$ for a two-phase autoresonant ion acoustic wave excited by two
driving counter propagating traveling waves with $k_{1}=-0.5$ and
$k_{2}=1$ obtained by solving the weakly nonlinear reduced dynamical
equations~(\ref{eq:dI1_dt})\textendash (\ref{eq:dPHI2_dt}). The
red vertical line indicates the termination of the drive at $\tau=12$.
The parameters used in the simulation are the same as in Figs.~\hyperref[fig_full]{\ref{fig_full}(a)},~\ref{fig_max_u},
and~\ref{fig_I_Phi_eps=00003D1.0}.}
\end{figure*}

\noindent 
\begin{figure}[t]
\includegraphics[width=1\columnwidth]{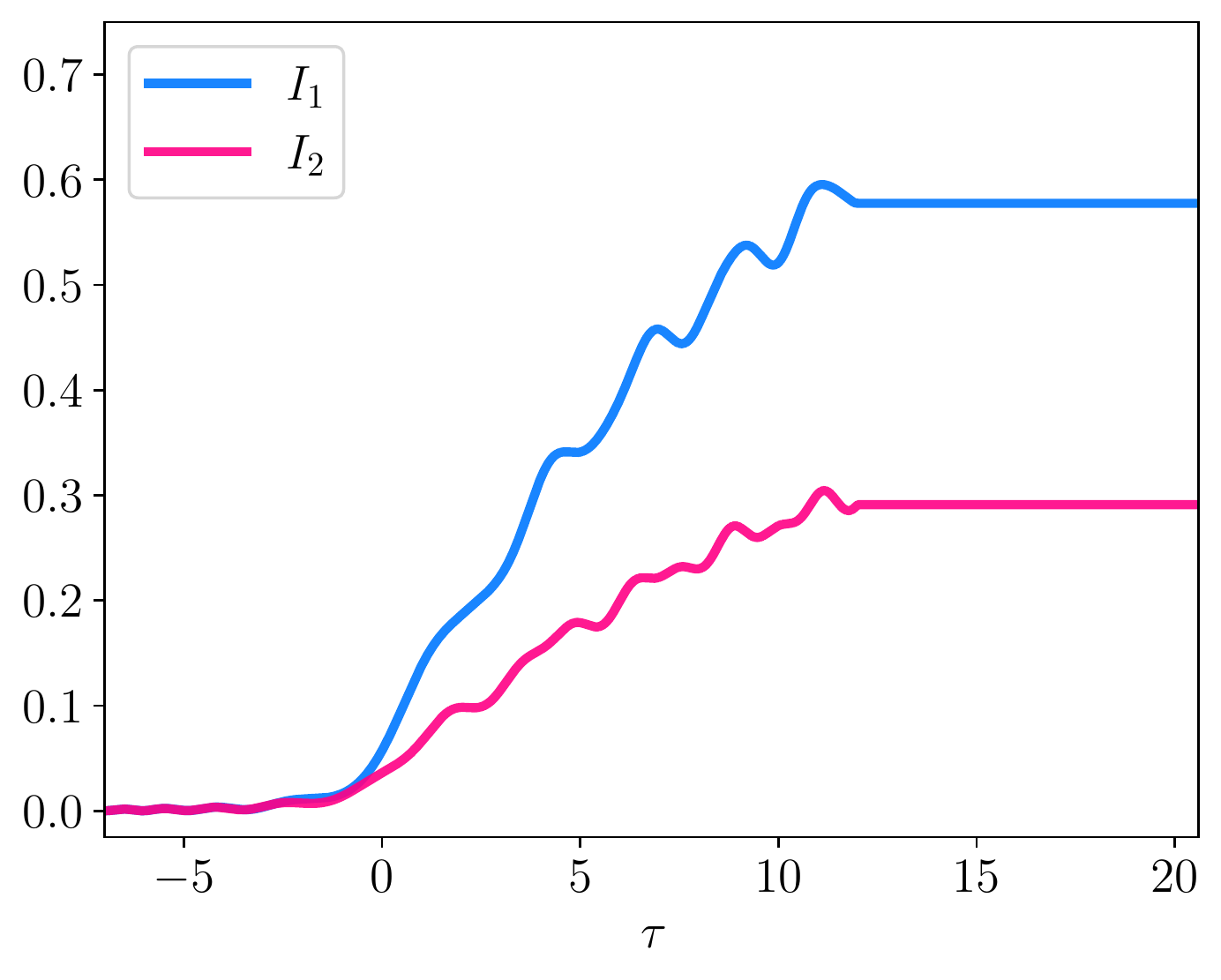}

\includegraphics[width=1\columnwidth]{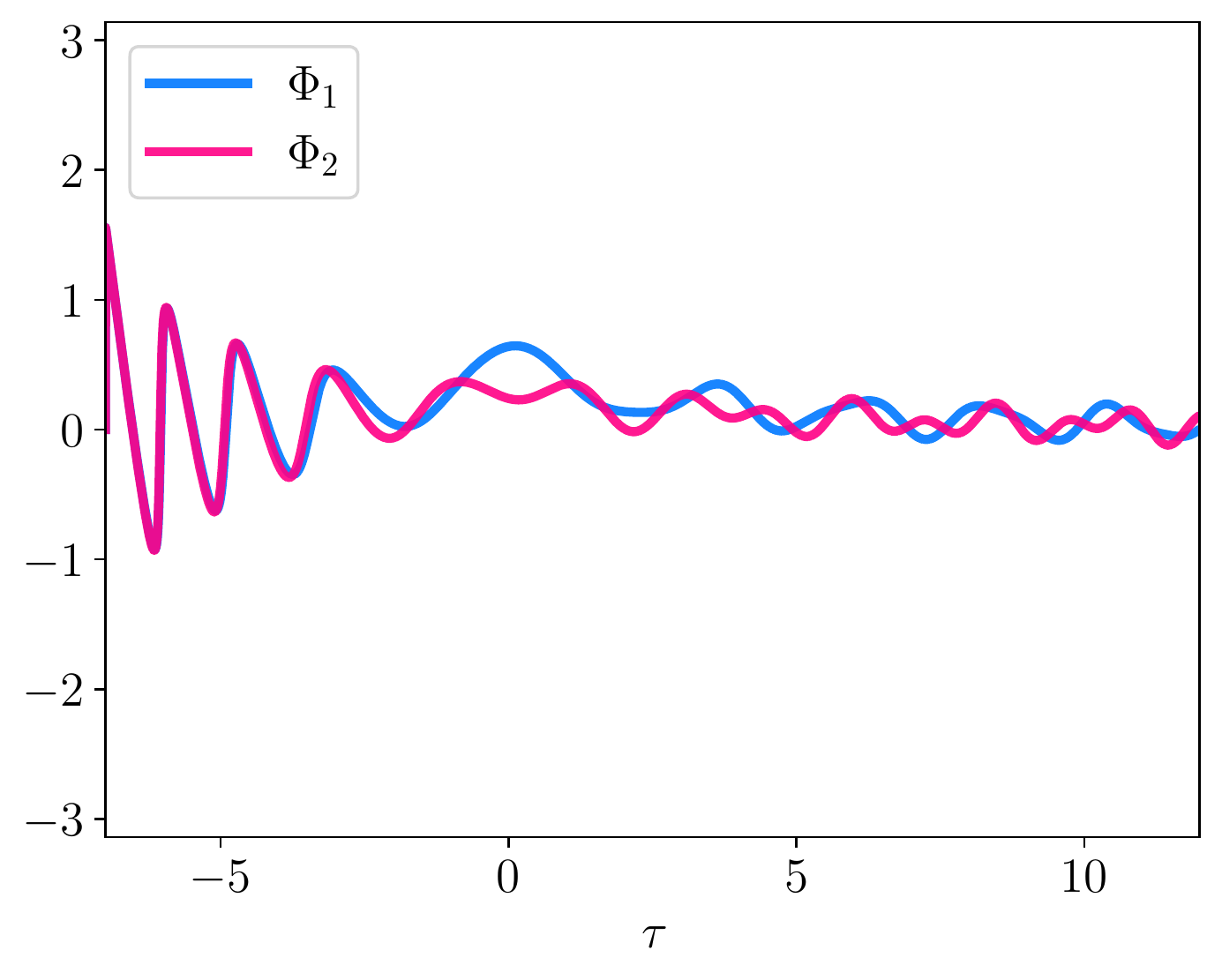}

\caption{\label{fig_I_Phi_eps=00003D1.0}The effective actions $I_{1},I_{2}$
(top subplot) and the phase mismatches $\Phi_{1},\Phi_{2}$ (bottom
subplot) versus slow time $\tau=\sqrt{\alpha_{1}}t$ obtained by solving
the weakly nonlinear reduced dynamical equations~(\ref{eq:dI1_dt})\textendash (\ref{eq:dPHI2_dt}).
The parameters used in the simulation are the same as in Figs.~\hyperref[fig_full]{\ref{fig_full}(a)},~\ref{fig_max_u},
and~\ref{fig_weakly}.}
\end{figure}

We start with a warm fluid model of ion acoustic waves in plasma described
by the following system of continuity, momentum, and Poisson's equations:

\begin{gather}
n_{t}+\left(nu\right)_{x}=0,\label{eq:n_t}\\
u_{t}+uu_{x}=-\varphi_{x}-\Delta^{2}nn_{x},\label{eq:u_t}\\
\varphi_{xx}=e^{\varphi+\varphi_{d}}-n.\label{eq:phi_xx}
\end{gather}

Here $n$ is the ion density, $u$ is the ion fluid velocity, $\Delta^{2}=3u_{th}^{2}$,
where $u_{th}$ is the ion thermal velocity, $\varphi$ is the electric
potential, and $\varphi_{d}$ is the driving potential. All variables
and parameters are dimensionless, such that the time is measured in
terms of the inverse ion plasma frequency $\omega_{pi}^{-1}=\sqrt{m_{i}/m_{e}}\omega_{p}^{-1}$,
the distance in terms of the Debye length $\lambda_{D}=u_{e}/\omega_{p}$,
and, consequently, the velocities are measured in terms of the modified
electron thermal velocity $\sqrt{m_{e}/m_{i}}u_{e}$. The plasma density
and the electric potential are normalized with respect to the unperturbed
plasma density and $k_{B}T_{e}/e$, respectively. The driving potential
consists of two small amplitude ($\varepsilon_{1}$, $\varepsilon_{2}$)
traveling waves and has the following form:

\begin{equation}
\varphi_{d}=\varepsilon_{1}\cos\left(\theta_{d,1}\right)+\varepsilon_{2}\cos\left(\theta_{d,2}\right),
\end{equation}

\noindent where the traveling wave drives have driving phases $\theta_{d,i}=k_{i}x-\int\omega_{d,i}\left(t\right)dt$
with slowly varying driving frequencies $\omega_{d,i}\left(t\right)=-d\theta_{d,i}/dt$
($i=1,2$).

We can solve the system of the nonlinear equations~(\ref{eq:n_t})\textendash (\ref{eq:phi_xx})
numerically using the water bag model method similar to the procedure
described in Refs.~\citep{Friedland2017,Friedland2019}.

To be specific, let us consider two driving counter propagating traveling
waves with $k_{1}=-0.5$ and $k_{2}=1$. We use linearly chirped driving
frequencies for $t\leq0$ and $\arctan$ drive for $t>0$ ($i=1,2$):

\begin{equation}
\omega_{d,i}=\begin{cases}
\omega_{a,i}+\alpha_{i}t, & t\leq0,\\
\omega_{a,i}+\alpha_{i}T_{i}\arctan\left(\frac{t}{T_{i}}\right), & t>0.
\end{cases}\label{eq:w_d}
\end{equation}

Here, $\omega_{a,i}\left(k_{i}\right)$ ($i=1,2$) are the frequencies
given by the linear ion acoustic wave dispersion relation:

\begin{equation}
\omega_{a,i}\left(k_{i}\right)=\left|k_{i}\right|\sqrt{\frac{1}{1+k_{i}^{2}}+\Delta^{2}},
\end{equation}

\noindent and we use equal chirp rates $\alpha_{1}=\alpha_{2}=2.5\times10^{-5}$
for $t\leq0$ and $T_{i}=2\Delta\omega_{i}/\pi\alpha_{i}$ ($i=1,2$),
$\Delta\omega_{1}=\Delta\omega_{2}=0.07$ for $t>0$. We also slowly
build up the driving amplitudes as $\varepsilon_{i}=\bar{\varepsilon}_{i}\left[0.5+\arctan\left(t/T_{i}\right)/\pi\right]$,
$\bar{\varepsilon}_{i}=16\alpha_{i}^{3/4}$ ($i=1,2$) to have a smoother
entrance into the autoresonant regime. The ion thermal velocity is
chosen as $u_{th}=0.003$.

The results of the numerical simulations are presented in Fig.~\ref{fig_full}.
Figure~\hyperref[fig_full]{\ref{fig_full}(a)} shows a colormap of
the electron density $n_{e}\left(x,\tau\right)$ approximated by $e^{\varphi\left(x,\tau\right)}$
as a function of $x$ and $\tau$, where we introduced a slow time
variable $\tau=\sqrt{\alpha_{1}}t$. We can clearly see in Fig.~\hyperref[fig_full]{\ref{fig_full}(a)}
a crystal-like quasiperiodic spatiotemporal structure representing
a large amplitude $\left(\delta n_{e}/n_{e}\sim1\right)$ two-phase
strongly nonlinear ion acoustic wave excited by the two driving counter
propagating traveling waves. Figures~\hyperref[fig_full]{\ref{fig_full}(b)}
and \hyperref[fig_full]{\ref{fig_full}(c)} show a colormap of $n_{e}\left(x,\tau\right)\approx e^{\varphi\left(x,\tau\right)}$
but when driven by just one of the driving components with $k_{1}=-0.5$
and $k_{2}=1$, respectively. We can see that an excitation of a single
phase large amplitude ion acoustic wave creates traveling wave-type
spatiotemporal photonic plasma structures. In these single phase solutions
the phases remain constant along the characteristic directions given
by the phase velocity of the driving waves. The same two characteristic
directions can also be seen in the two-phase wave solution in Fig.~\hyperref[fig_full]{\ref{fig_full}(a)}.
We start our simulations at $\tau=-7$ and stop the driving at $\tau=12$
(which is indicated by the red vertical line in Fig.~\ref{fig_full}.
At $\tau=0$ both driving waves pass the linear resonances and then
the system enters the autoresonant regime and its excitation amplitude
increases to preserve the resonances with the drives. This can be
seen in Fig.~\ref{fig_max_u}, which shows the maximum value over
$x$ of the ion fluid velocity $u\left(x,\tau\right)$ versus slow
time $\tau=\sqrt{\alpha_{1}}t$. Note that the crystal-like structure
is preserved in time after we turn off the drive at $\tau=12$, suggesting
formation of some fundamental mode of the system.

The limiting factor as to what amplitudes we can excite the ion acoustic
waves is determined by the kinetic wave breaking \citep{Kakad2017,Riconda2005,Friedland2017,Friedland2019}.
For the cold plasma this occurs when the ion fluid velocity exceeds
the absolute value of the phase velocity of at least one of the driving
waves. As we can see in Fig.~\ref{fig_max_u}, the maximum amplitude
of the ion fluid velocity $u$ stays below the absolute value of the
phase velocities of the driving waves; we thereby avoid the wave-breaking
limit for the parameters chosen in our example.

In the next section, we are going to cast the problem into the Lagrangian
form and develop an adiabatic, weakly nonlinear theory using Whitham's
averaged variational principle \citep{Whitham1999}. These analytical
results will allow us to understand how to control and choose parameters
for the excitation of large amplitude multiphase ion acoustic waves.

\section{\label{sec_III}The Lagrangian formulation and Whitham\textquoteright s
variational principle}

For convenience, let us introduce two new potentials $\psi$ and $\sigma$
via $u=\psi_{x}$, $n=1+\sigma_{x}$. The system of the nonlinear
equations~(\ref{eq:n_t})\textendash (\ref{eq:phi_xx}) in terms
of the new variables is then

\begin{gather}
\sigma_{xt}+\left[\left(1+\sigma_{x}\right)\psi_{x}\right]_{x}=0,\label{eq:sigma_xt}\\
\psi_{xt}+\psi_{x}\psi_{xx}=-\varphi_{x}-\Delta^{2}\left(1+\sigma_{x}\right)\sigma_{xx},\label{eq:psi_xt}\\
\varphi_{xx}\approx\left(1+\varphi_{d}\right)e^{\varphi}-\sigma_{x}-1,\label{eq:phi_xx_2}
\end{gather}

\noindent where we assumed that the drive is small: $e^{\varphi_{d}}\approx1+\varphi_{d}$.

Equations~(\ref{eq:sigma_xt})\textendash (\ref{eq:phi_xx_2}) can
be derived from the Lagrangian variational principle $\delta\left(\int Ldxdt\right)=0$
with the corresponding Lagrangian density $L$ given by

\begin{multline}
L=\frac{1}{2}\varphi_{x}^{2}+V\left(\varphi\right)-\frac{1}{2}\left(\psi_{t}\sigma_{x}+\psi_{x}\sigma_{t}\right)\\
-\left(\frac{1}{2}\psi_{x}^{2}+\varphi\right)\left(1+\sigma_{x}\right)-\frac{\Delta^{2}}{2}\sigma_{x}^{2}\left(1+\frac{1}{3}\sigma_{x}\right)+\varphi_{d}\varphi,\label{eq:Lagrangian}
\end{multline}

\noindent where $V\left(\varphi\right)=\varphi+\frac{1}{2}\varphi^{2}+\frac{1}{6}\varphi^{3}+\frac{1}{24}\varphi^{4}$.

The Lagrangian form of the problem together with the slow adiabatic
synchronization (autoresonance) procedure we employ to excite multiphase
waves suggest that it should be possible to use Whitham\textquoteright s
averaged variational principle \citep{Whitham1999} to derive weakly
nonlinear analytical results for our problem. 

We proceed by writing the following ansatz describing the two-phase
{[}$\theta_{1}=k_{1}x-\int\omega_{1}\left(t\right)dt$, $\theta_{2}=k_{2}x-\int\omega_{2}\left(t\right)dt${]}
solutions for the potentials $\sigma$, $\psi$, $\varphi$:

\begin{multline}
\sigma=\tilde{A}_{10}\sin\left(\theta_{1}\right)+\tilde{A}_{01}\sin\left(\theta_{2}\right)\\
+\tilde{A}_{11}\sin\left(\theta_{1}+\theta_{2}\right)+\tilde{A}_{1,-1}\sin\left(\theta_{1}-\theta_{2}\right)\\
+\tilde{A}_{20}\sin\left(2\theta_{1}\right)+\tilde{A}_{02}\sin\left(2\theta_{2}\right),\label{eq:sigma_ansatz}
\end{multline}

\begin{multline}
\psi=\tilde{B}_{10}\sin\left(\theta_{1}\right)+\tilde{B}_{01}\sin\left(\theta_{2}\right)\\
+\tilde{B}_{11}\sin\left(\theta_{1}+\theta_{2}\right)+\tilde{B}_{1,-1}\sin\left(\theta_{1}-\theta_{2}\right)\\
+\tilde{B}_{20}\sin\left(2\theta_{1}\right)+\tilde{B}_{02}\sin\left(2\theta_{2}\right),\label{eq:psi_ansatz}
\end{multline}

\begin{multline}
\varphi=C_{00}+C_{10}\cos\left(\theta_{1}\right)+C_{01}\cos\left(\theta_{2}\right)\\
+C_{11}\cos\left(\theta_{1}+\theta_{2}\right)+C_{1,-1}\cos\left(\theta_{1}-\theta_{2}\right)\\
+C_{20}\cos\left(2\theta_{1}\right)+C_{02}\cos\left(2\theta_{2}\right).\label{eq:phi_ansatz}
\end{multline}

Here, we view the amplitudes $\tilde{A}_{10}$, $\tilde{A}_{01}$,
$\tilde{B}_{10}$, $\tilde{B}_{01}$, $C_{10}$, $C_{01}$ as the
first-order coefficients, while the amplitudes $\tilde{A}_{20}$,
$\tilde{A}_{02}$, $\tilde{B}_{20}$, $\tilde{B}_{02}$, $C_{20}$,
$C_{02}$, $\tilde{A}_{11}$, $\tilde{A}_{1,-1}$, $\tilde{B}_{11}$,
$\tilde{B}_{1,-1}$, $C_{11}$, $C_{1,-1}$, and $C_{00}$ as the
second-order coefficients. It is also convenient to write the driving
potential $\varphi_{d}$ in the form with explicit phase mismatches
$\Phi_{i}=\theta_{i}-\theta_{d,i}$ ($i=1,2$) between phases of the
solutions $\theta_{1}$, $\theta_{2}$ and the driving phases $\theta_{d,1}$,
$\theta_{d,2}$:

\begin{equation}
\varphi_{d}=\varepsilon_{1}\cos\left(\theta_{1}-\Phi_{1}\right)+\varepsilon_{2}\cos\left(\theta_{2}-\Phi_{2}\right).
\end{equation}

Furthermore, here and in the following it is assumed that the phases
$\theta_{1}$, $\theta_{2}$ are varying rapidly, while all the coefficients
in our ansatz and $\Phi_{1}$, $\Phi_{2}$ are slow functions of time.

In the linear case without phase mismatches we have the following
solutions:

\begin{gather}
\sigma=\tilde{A}_{10}\sin\left(\theta_{1}\right)+\tilde{A}_{01}\sin\left(\theta_{2}\right),\\
\psi=\tilde{B}_{10}\sin\left(\theta_{1}\right)+\tilde{B}_{01}\sin\left(\theta_{2}\right),\\
\varphi=C_{10}\cos\left(\theta_{1}\right)+C_{01}\cos\left(\theta_{2}\right),
\end{gather}

\noindent where the amplitudes $C_{10}$ and $C_{01}$ satisfy

\begin{align}
C_{10}\left(\frac{k_{1}^{2}}{\omega_{1}^{2}-\Delta^{2}k_{1}^{2}}-1-k_{1}^{2}\right)=\varepsilon_{1},\\
C_{01}\left(\frac{k_{2}^{2}}{\omega_{2}^{2}-\Delta^{2}k_{2}^{2}}-1-k_{2}^{2}\right)=\varepsilon_{2},
\end{align}

\noindent and the amplitudes $\tilde{A}_{10}$, $\tilde{A}_{01}$,
$\tilde{B}_{10}$, $\tilde{B}_{01}$ are expressed through $C_{10}$
and $C_{01}$ as follows:

\begin{align}
\tilde{A}_{10} & =\frac{k_{1}}{\omega_{1}^{2}-\Delta^{2}k_{1}^{2}}C_{10},\label{eq:A_lin_1}\\
\tilde{A}_{01} & =\frac{k_{2}}{\omega_{2}^{2}-\Delta^{2}k_{2}^{2}}C_{01},\label{eq:A_lin_2}\\
\tilde{B}_{10} & =\frac{\omega_{1}}{\omega_{1}^{2}-\Delta^{2}k_{1}^{2}}C_{10},\label{eq:B_lin_1}\\
\tilde{B}_{01} & =\frac{\omega_{2}}{\omega_{2}^{2}-\Delta^{2}k_{2}^{2}}C_{01}.\label{eq:B_lin_2}
\end{align}

The next crucial step is to find the averaged Lagrangian density over
the rapidly varying phases $\theta_{1}$, $\theta_{2}$:
\begin{equation}
\bar{L}=\left\langle L\left(\theta_{1},\theta_{2},t\right)\right\rangle _{\theta_{1},\theta_{2}}=\int L\left(\theta_{1},\theta_{2},t\right)\frac{d\theta_{1}}{2\pi}\frac{d\theta_{2}}{2\pi},
\end{equation}

\noindent which will depend only on slow variables: the amplitudes
of various harmonics and the phase mismatches.

After long but straightforward calculations one can obtain the average
of Eq.~(\ref{eq:Lagrangian}) over the rapidly varying phases. This
derivation of the averaged Lagrangian can be found in Appendix~\ref{Appendix_A}.

\noindent 
\begin{figure}[t]
\includegraphics[width=1\columnwidth]{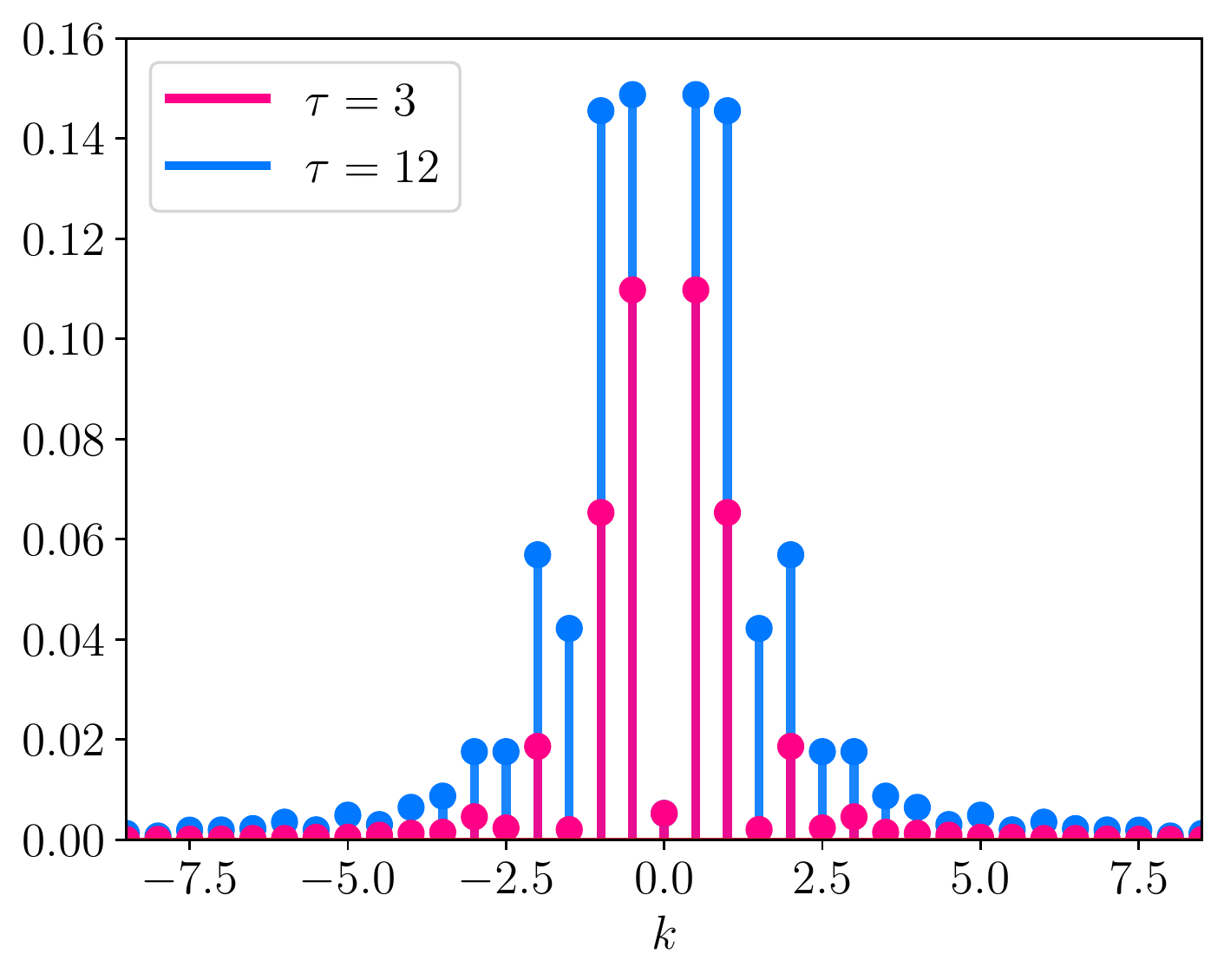}

\includegraphics[width=1\columnwidth]{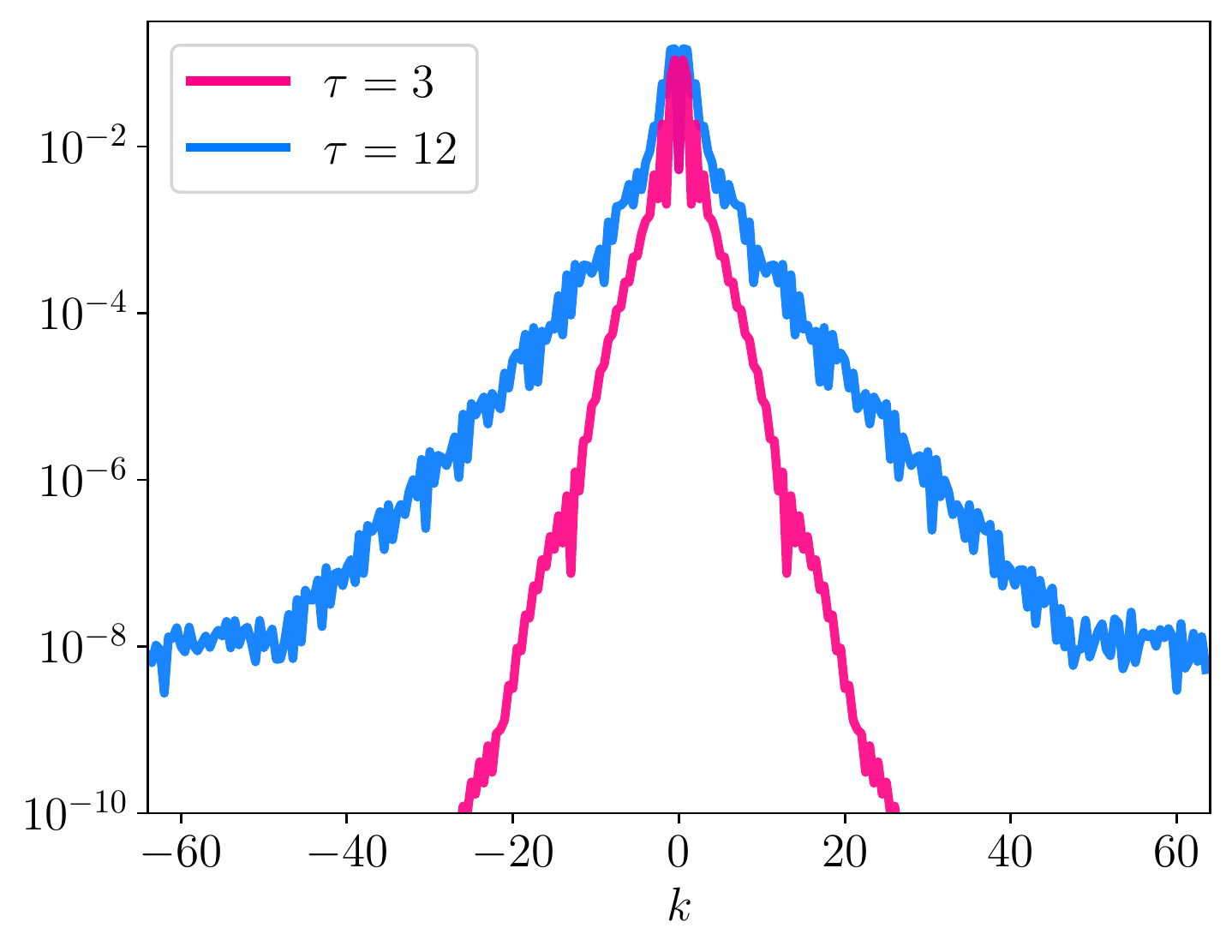}

\caption{\label{fig_spec}The spectrum of the harmonics of the ion fluid velocity
$u\left(x,\tau\right)$ in $k$-space obtained by solving the fully
nonlinear equations~(\ref{eq:n_t})\textendash (\ref{eq:phi_xx})
at $\tau=3$ (pink line) and $\tau=12$ (blue line) in linear (top
subplot) and logarithmic (bottom subplot) scales. The parameters used
in the simulations are the same as in Figs.~\hyperref[fig_full]{\ref{fig_full}(a)},~\ref{fig_max_u},~\ref{fig_weakly},
and~\ref{fig_I_Phi_eps=00003D1.0}.}
\end{figure}

\subsection*{Weakly nonlinear equations}

Now, having obtained the averaged Lagrangian, we can derive the weakly
nonlinear equations that describe the evolution of the wave amplitude
by applying Whitham\textquoteright s variational procedure \citep{Whitham1999}.

First, we take variations with respect to the phases:

\begin{align}
\frac{d}{dt}\left(\frac{\partial\bar{L}}{\partial\dot{\theta}_{1}}\right)-\frac{\partial\bar{L}}{\partial\theta_{1}} & =-\frac{d}{dt}\left(\frac{\partial\bar{L}}{\partial\omega_{1}}\right)-\frac{\partial\bar{L}}{\partial\Phi_{1}}=0,\\
\frac{d}{dt}\left(\frac{\partial\bar{L}}{\partial\dot{\theta}_{2}}\right)-\frac{\partial\bar{L}}{\partial\theta_{2}} & =-\frac{d}{dt}\left(\frac{\partial\bar{L}}{\partial\omega_{2}}\right)-\frac{\partial\bar{L}}{\partial\Phi_{2}}=0.
\end{align}

Keeping the lowest significant order terms and using the linear relations
(\ref{eq:A_lin_1})\textendash (\ref{eq:B_lin_2}), we get

\begin{alignat}{1}
\frac{d}{dt}\left[\frac{\omega_{1}k_{1}^{2}}{\left(\omega_{1}^{2}-\Delta^{2}k_{1}^{2}\right)^{2}}C_{10}^{2}\right] & =\varepsilon_{1}C_{10}\sin\left(\Phi_{1}\right),\label{eq:dC10/dt}\\
\frac{d}{dt}\left[\frac{\omega_{2}k_{2}^{2}}{\left(\omega_{2}^{2}-\Delta^{2}k_{2}^{2}\right)^{2}}C_{01}^{2}\right] & =\varepsilon_{2}C_{01}\sin\left(\Phi_{2}\right).\label{eq:dC01/dt}
\end{alignat}

Likewise, we can obtain from the variations with respect to the first-order
amplitudes {[}see Eqs.~(\ref{eq:C10})\textendash (\ref{eq:A01})
in Appendix~\ref{Appendix_B}{]}:

\begin{multline}
\left(\frac{k_{1}^{2}}{\omega_{1}^{2}-\Delta^{2}k_{1}^{2}}-1-k_{1}^{2}\right)C_{10}=C_{10}^{2}P\left(k_{1},\omega_{1}\right)C_{10}\\
+C_{01}C_{10}Q\left(k_{1},\omega_{1};k_{2},\omega_{2}\right)C_{01}+\varepsilon_{1}\cos\left(\Phi_{1}\right),\label{eq:PQ_w1}
\end{multline}

\begin{multline}
\left(\frac{k_{2}^{2}}{\omega_{2}^{2}-\Delta^{2}k_{2}^{2}}-1-k_{2}^{2}\right)C_{01}=C_{01}^{2}P\left(k_{2},\omega_{2}\right)C_{01}\\
+C_{01}C_{10}Q\left(k_{1},\omega_{1};k_{2},\omega_{2}\right)C_{10}+\varepsilon_{2}\cos\left(\Phi_{2}\right),\label{eq:PQ_w2}
\end{multline}

\noindent where the functions $P\left(k_{1},\omega_{1}\right)$ and
$Q\left(k_{1},\omega_{1};k_{2},\omega_{2}\right)$ are defined in
Appendix~\ref{Appendix_C}.

Expanding around the linear ion acoustic dispersion relation $\omega_{i}=\omega_{a,i}+\Delta\omega_{i}$
($i=1,2$), we get from Eqs.~(\ref{eq:PQ_w1}) and (\ref{eq:PQ_w2})
the following expressions:

\begin{multline}
\Delta\omega_{1}=-\frac{\left(\omega_{1}^{2}-\Delta^{2}k_{1}^{2}\right)^{2}}{2\omega_{1}k_{1}^{2}}P\left(k_{1},\omega_{1}\right)C_{10}^{2}\\
-\frac{\left(\omega_{1}^{2}-\Delta^{2}k_{1}^{2}\right)^{2}}{2\omega_{1}k_{1}^{2}}Q\left(k_{1},\omega_{1};k_{2},\omega_{2}\right)C_{01}^{2}\\
-\frac{\left(\omega_{1}^{2}-\Delta^{2}k_{1}^{2}\right)^{2}}{2\omega_{1}k_{1}^{2}}\frac{\varepsilon_{1}}{C_{10}}\cos\left(\Phi_{1}\right),\label{eq:dw1}
\end{multline}

\begin{multline}
\Delta\omega_{2}=-\frac{\left(\omega_{2}^{2}-\Delta^{2}k_{2}^{2}\right)^{2}}{2\omega_{2}k_{2}^{2}}P\left(k_{2},\omega_{2}\right)C_{01}^{2}\\
-\frac{\left(\omega_{2}^{2}-\Delta^{2}k_{2}^{2}\right)^{2}}{2\omega_{2}k_{2}^{2}}Q\left(k_{1},\omega_{1};k_{2},\omega_{2}\right)C_{10}^{2}\\
-\frac{\left(\omega_{2}^{2}-\Delta^{2}k_{2}^{2}\right)^{2}}{2\omega_{2}k_{2}^{2}}\frac{\varepsilon_{2}}{C_{01}}\cos\left(\Phi_{2}\right).\label{eq:dw2}
\end{multline}

The above expressions~(\ref{eq:dw1}) and~(\ref{eq:dw2}) show that
due to the nonlinear nature of the system, the waves acquire frequency
shifts (the first two terms), which can be adjusted to the chirped
driving frequencies continuously (see the last term), yielding control
of the wave amplitudes.

Assuming linear driving frequency chirps $\omega_{d,i}=\omega_{a,i}+\alpha_{i}t$
($i=1,2$) and defining

\begin{gather}
I_{1}=\frac{2\omega_{1}k_{1}^{2}}{\left(\omega_{1}^{2}-\Delta^{2}k_{1}^{2}\right)^{2}}C_{10}^{2},\;I_{2}=\frac{2\omega_{2}k_{2}^{2}}{\left(\omega_{2}^{2}-\Delta^{2}k_{2}^{2}\right)^{2}}C_{01}^{2},\\
a=\frac{\left(\omega_{1}^{2}-\Delta^{2}k_{1}^{2}\right)^{4}}{4\omega_{1}^{2}k_{1}^{4}}P\left(k_{1},\omega_{1}\right),\\
b=\frac{\left(\omega_{1}^{2}-\Delta^{2}k_{1}^{2}\right)^{2}}{2\omega_{1}k_{1}^{2}}\frac{\left(\omega_{2}^{2}-\Delta^{2}k_{2}^{2}\right)^{2}}{2\omega_{2}k_{2}^{2}}Q\left(k_{1},\omega_{1},k_{2},\omega_{2}\right),\\
c=\frac{\left(\omega_{2}^{2}-\Delta^{2}k_{2}^{2}\right)^{4}}{4\omega_{2}^{2}k_{2}^{4}}P\left(k_{2},\omega_{2}\right),\\
\epsilon_{1}=\frac{2\left(\omega_{1}^{2}-\Delta^{2}k_{1}^{2}\right)}{\left|k_{1}\right|}\varepsilon_{1},\;\epsilon_{2}=\frac{2\left(\omega_{2}^{2}-\Delta^{2}k_{2}^{2}\right)}{\left|k_{2}\right|}\varepsilon_{2},
\end{gather}

\noindent we can rewrite Eqs.~(\ref{eq:dC10/dt}), (\ref{eq:dC01/dt}),
(\ref{eq:dw1}), and~(\ref{eq:dw2}) to obtain the following system
of weakly nonlinear evolution equations:

\begin{alignat}{1}
\frac{dI_{1}}{dt} & =\epsilon_{1}\sqrt{\frac{I_{1}}{2\omega_{1}}}\sin\left(\Phi_{1}\right),\label{eq:dI1_dt}\\
\frac{dI_{2}}{dt} & =\epsilon_{2}\sqrt{\frac{I_{2}}{2\omega_{2}}}\sin\left(\Phi_{2}\right),\label{eq:dI2_dt}\\
\frac{d\Phi_{1}}{dt} & =aI_{1}+bI_{2}+\alpha_{1}t+\frac{\epsilon_{1}}{2\sqrt{2\omega_{1}I_{1}}}\cos\left(\Phi_{1}\right),\label{eq:dPHI1_dt}\\
\frac{d\Phi_{2}}{dt} & =bI_{1}+cI_{2}+\alpha_{2}t+\frac{\epsilon_{2}}{2\sqrt{2\omega_{2}I_{2}}}\cos\left(\Phi_{2}\right).\label{eq:dPHI2_dt}
\end{alignat}

Here, in principle, the linear frequency chirps $\alpha_{i}t$ ($i=1,2$)
can be replaced by any other functions of time as long as these functions
are sufficiently slow. In fact, we use the $\arctan$ frequency chirp
drive defined in Eq.~(\ref{eq:w_d}) in our simulations.

The numerical solution of the weakly nonlinear system~(\ref{eq:dI1_dt})\textendash (\ref{eq:dPHI2_dt})
is presented in Figs.~\ref{fig_max_u}\textendash \ref{fig_I_Phi_eps=00003D1.0}.
We use the drive and other parameters identical to those used in the
numerical solution of the fully nonlinear system described in the
previous section, which is presented in Figs.~\hyperref[fig_full]{\ref{fig_full}(a)}
and~\ref{fig_max_u}. As can be seen in Fig.~\ref{fig_max_u} and
by comparison between Figs.~\hyperref[fig_full]{\ref{fig_full}(a)}
and~\ref{fig_weakly}, the analytically derived weakly nonlinear
system is indeed a good approximation of the fully nonlinear problem.
We further observe in Fig.~\ref{fig_max_u} the absence of the low
frequency modulation in the weakly nonlinear solution after we stop
driving at $\tau=12$. The reason is that in this case $I_{1},I_{2}=\textrm{const}$,
as evident from Eqs.~(\ref{eq:dI1_dt})\textendash (\ref{eq:dI2_dt}),
which implies that all the slowly evolving amplitudes are constant
as well, while Eqs.~(\ref{eq:sigma_ansatz})\textendash (\ref{eq:phi_ansatz})
show that it is the slowly evolving amplitudes that are directly responsible
for the low frequency modulation.

The nature of the autoresonant excitation and phase locking is demonstrated
in Fig.~\ref{fig_I_Phi_eps=00003D1.0}, which shows the effective
actions $I_{1},I_{2}$ (top subplot) and the phase mismatches $\Phi_{1},\Phi_{2}$
(bottom subplot) as functions of slow time $\tau=\sqrt{\alpha_{1}}t$.
We can clearly see that as the system passes the linear resonance
at $\tau=0$, the phase mismatches are locked around zero, while the
effective actions $I_{1},I_{2}$ both enter the autoresonant regime
and grow in amplitude. At $\tau=12$ we turn off the drives and so
the effective actions remain constant. In the absence of the drives,
the phase mismatches are not meaningfully defined, therefore $\Phi_{1},\Phi_{2}$
are shown in the figures only until $\tau=12$. 

We can explicitly test whether our assumption regarding the form of
the ansatz used {[}see Eqs.~(\ref{eq:sigma_ansatz})\textendash (\ref{eq:phi_ansatz}){]}
is justified by performing the spectral analysis. Figure~\ref{fig_spec}
shows the spectral distribution of the harmonics of the ion fluid
velocity $u\left(x,\tau\right)$ in $k$-space at two moments of time:
$\tau=3$ (pink line) and $\tau=12$ (blue line), both in linear scale
(top subplot) and in logarithmic scale (bottom subplot). We can see
that the spectrum falls off dramatically with the increase in $\left|k\right|$;
only a handful of harmonics have a noticeable weight and can be considered
excited. At $\tau=3$ the excited harmonics correspond to the ones
used in the ansatz, so we see a very good agreement between the weakly
nonlinear and the fully nonlinear solutions, as evident in Fig.~\ref{fig_max_u}.
At $\tau=12$ we can see that harmonics with $\left|k\right|=2$ and
$\left|k\right|=2.5$, which are beyond the ones used in the ansatz,
acquire some small but not non-negligible weights, so the agreement
between the weakly nonlinear theory utilizing the ansatz given by
Eqs.~(\ref{eq:sigma_ansatz})\textendash (\ref{eq:phi_ansatz}) and
the fully nonlinear theory is not as good as at $\tau=3$, though
it is still decent. Thus, by checking the spectrum of the solutions,
one can verify whether the ansatz is adequate for the parameters used
in the simulations, and if necessary the ansatz can be extended to
include more harmonics and more precise weakly analytical theory can
be developed.

\noindent 
\begin{figure}[t]
\includegraphics[width=1\columnwidth]{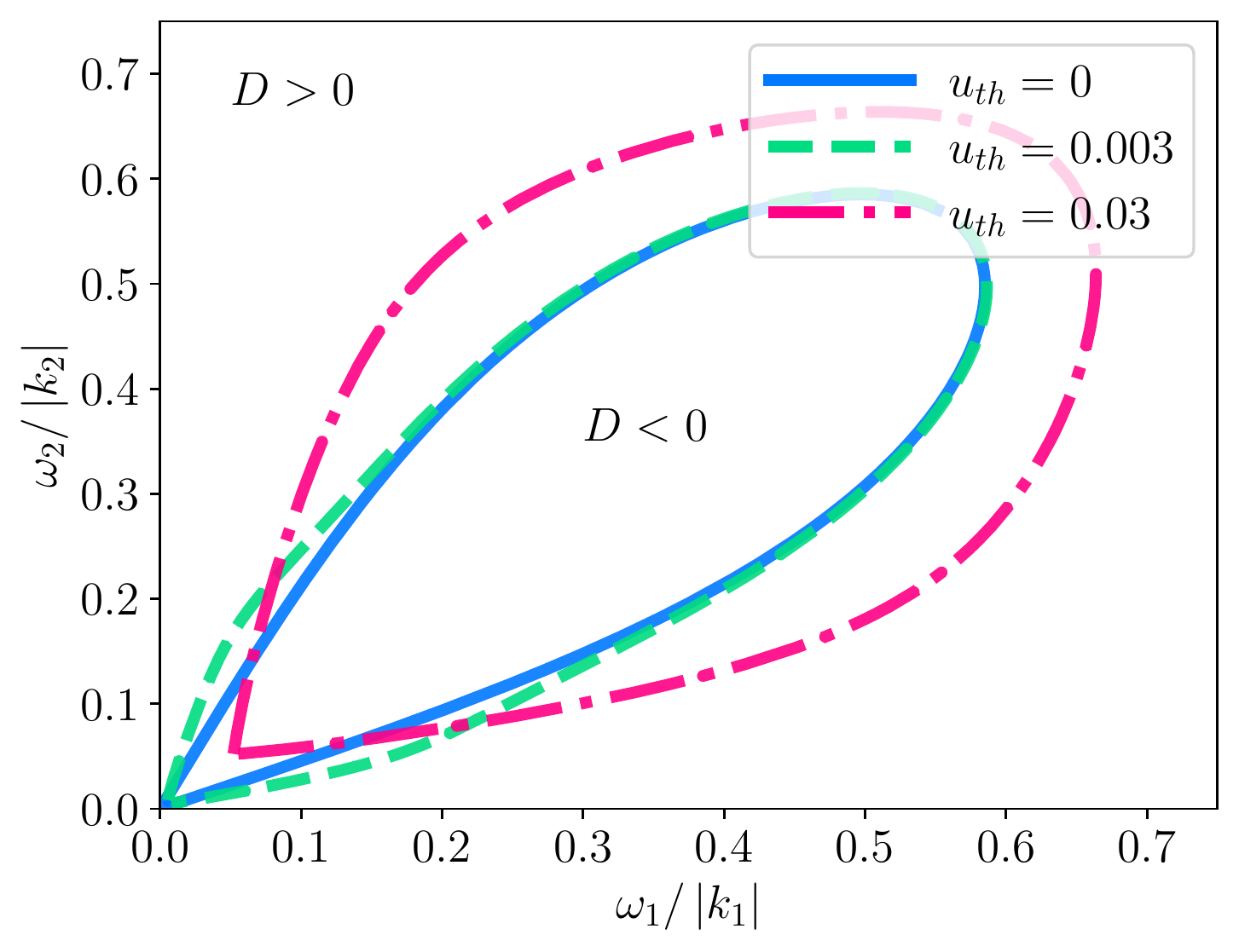}

\caption{\label{fig_D_vph_plane}The curves $D\left(k_{1},k_{2}\right)=0$
for three values of the ion thermal velocity: $u_{th}=0$ (solid blue
line), $u_{th}=0.003$ (dashed green line), and $u_{th}=0.03$ (dash-dotted
pink line), plotted in the $\left(\omega_{1}/\left|k_{1}\right|,\omega_{2}/\left|k_{2}\right|\right)$
plane assuming $k_{1}k_{2}>0$. $D$ is negative inside the curve
$D=0$ and positive outside the curve. For $k_{1}k_{2}<0$, $D$ is
always positive. The double autoresonance is possible only in the
regions where $D>0$.}
\end{figure}

\section{\label{sec:IV}The conditions for double autoresonance}

Equations~(\ref{eq:dI1_dt})\textendash (\ref{eq:dPHI2_dt}) have
the same form as a weakly nonlinear system studied in Ref.~\citep{Barth2007}.
Such a system is described by the following Hamiltonian for the effective
action ($I_{1}$, $I_{2}$) and angle ($\Phi_{1}$, $\Phi_{2}$) variables:

\begin{equation}
H=\omega_{1}I_{1}+\omega_{2}I_{2}+\frac{1}{2}aI_{1}^{2}+bI_{1}I_{2}+\frac{1}{2}cI_{2}^{2}+f\left(I_{1},I_{2},\Phi_{1},\Phi_{2},t\right),
\end{equation}

\noindent where

\begin{equation}
f\left(I_{1},I_{2},\Phi_{1},\Phi_{2},t\right)=\epsilon_{1}\sqrt{\frac{I_{1}}{2\omega_{1}}}\cos\left(\Phi_{1}\right)+\epsilon_{2}\sqrt{\frac{I_{2}}{2\omega_{2}}}\cos\left(\Phi_{2}\right).
\end{equation}

As shown in Ref.~\citep{Barth2007}, the possibility of double autoresonance
is determined by the signs of $D=ac-b^{2}$ and $\alpha_{1}\alpha_{2}$:
for the double autoresonance to occur, they must have the same sign.
From Eqs.~(\ref{eq:dI1_dt})\textendash (\ref{eq:dPHI2_dt}), we
see that in the case of the double autoresonance the asymptotic large
$t$ solutions for the actions are given by

\begin{equation}
\bar{I}_{1}=\frac{b\alpha_{2}-c\alpha_{1}}{D}t,\:\bar{I}_{2}=\frac{b\alpha_{1}-a\alpha_{2}}{D}t.
\end{equation}

Thus, in addition, for the double autoresonance to actually happen
the asymptotic actions $\bar{I}_{1}$, $\bar{I}_{2}$ defined above
must be positive.

There are two possible situations, depending on whether $k_{1}$ and
$k_{2}$ have the same sign. (1) If $k_{1}k_{2}>0$, we have $b<0$,
$a<0$, $c<0$. In this case $D$ can have both positive and negative
values. Figure~\ref{fig_D_vph_plane} shows the lines for $D=0$
in the plane formed by the absolute values of the phase velocities
of the driving waves $\left(\omega_{1}/\left|k_{1}\right|,\omega_{2}/\left|k_{2}\right|\right)$
for different values of the ion thermal velocity $u_{th}$. We can
see that the ion thermal velocity $u_{th}$, though not an extremely
sensitive parameter, nevertheless determines the regions of positive
and negative values of $D$. Thus the cold ion model should be used
carefully and the thermal ion velocity should in general be taken
into account. In the region where $D>0$, for the double autoresonance
to occur $\alpha_{1}\alpha_{2}$ must be positive. In addition, for
positive $\bar{I}_{1}$, $\bar{I}_{2}$ we must have $\left|b\right|/\left|a\right|<\alpha_{2}/\alpha_{1}<\left|c\right|/\left|b\right|$.
In contrast, if $D<0$, we must have $\alpha_{1}\alpha_{2}<0$. However,
in this case it is impossible for both $\bar{I}_{1}$ and $\bar{I}_{2}$
to be positive at the same time. Thus, if $k_{1}k_{2}>0$, the double
autoresonance is possible only in the region where $D>0$; in this
region we also must have $\alpha_{1}\alpha_{2}>0$ and $\left|b\right|/\left|a\right|<\alpha_{2}/\alpha_{1}<\left|c\right|/\left|b\right|$.
(2) The second possibility is $k_{1}k_{2}<0$, then we have $b>0$,
$a<0$, $c<0$. In this case $D$ is always positive, and, consequently,
for double autoresonance we must have $\alpha_{1}\alpha_{2}>0$. In
addition, to have positive $\bar{I}_{1}$, $\bar{I}_{2}$ only the
case when both $\alpha_{1},\alpha_{2}$ are positive works. Thus,
in the case where $k_{1}k_{2}<0$, the double autoresonance is possible
when $\alpha_{1},\alpha_{2}>0$. We do not deal with the degenerate
case of $k_{1}=k_{2}$ in this paper.

\noindent 
\begin{figure}
\includegraphics[width=1\columnwidth]{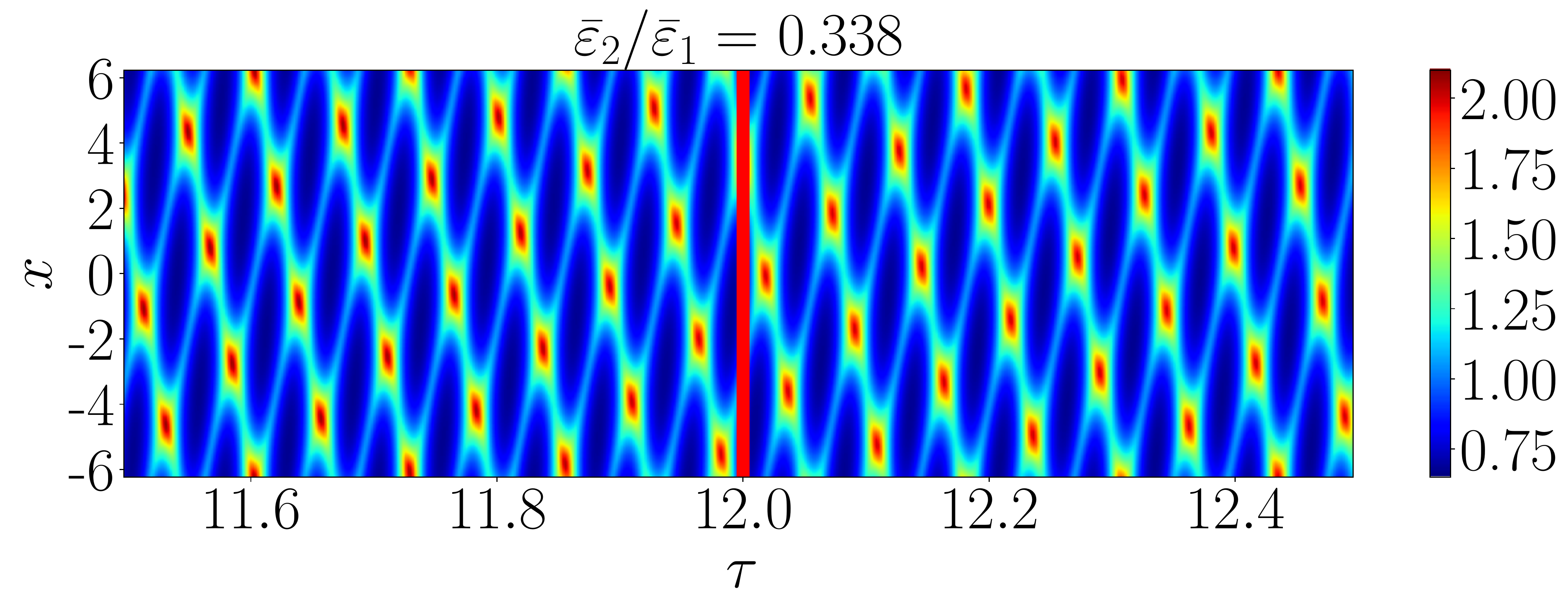}

\includegraphics[width=1\columnwidth]{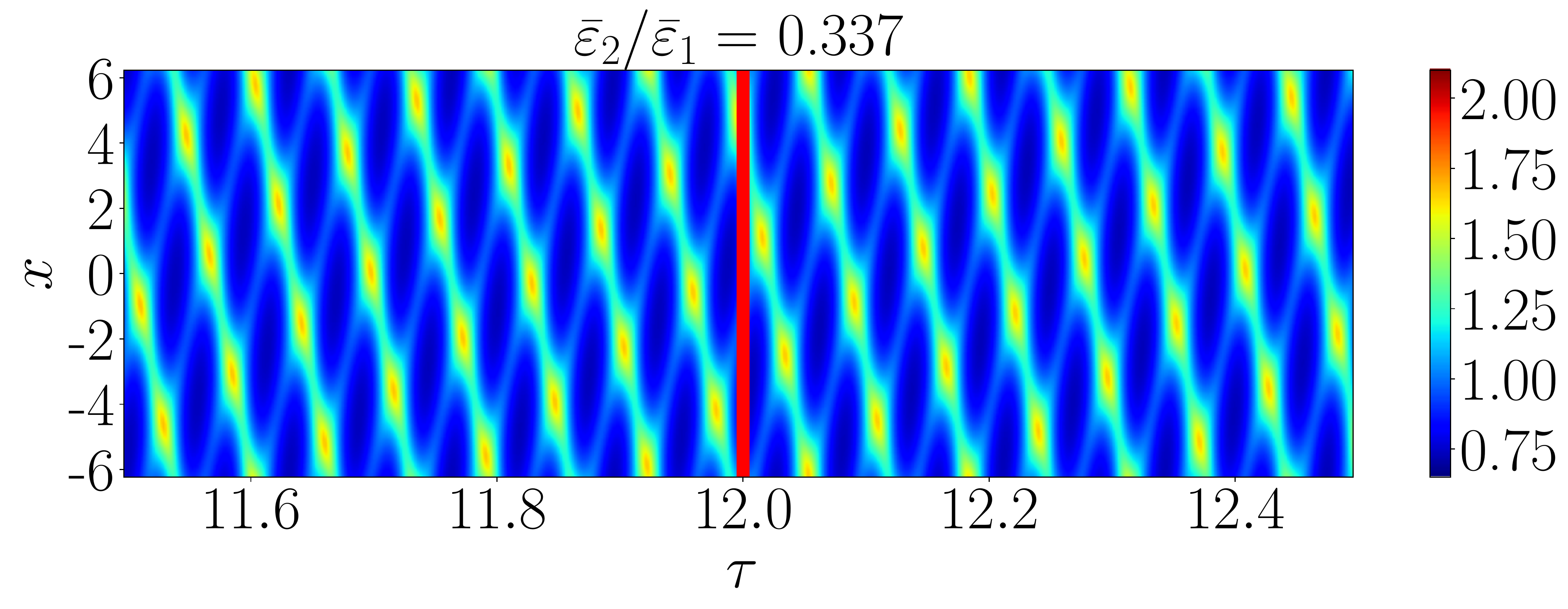}

\includegraphics[width=1\columnwidth]{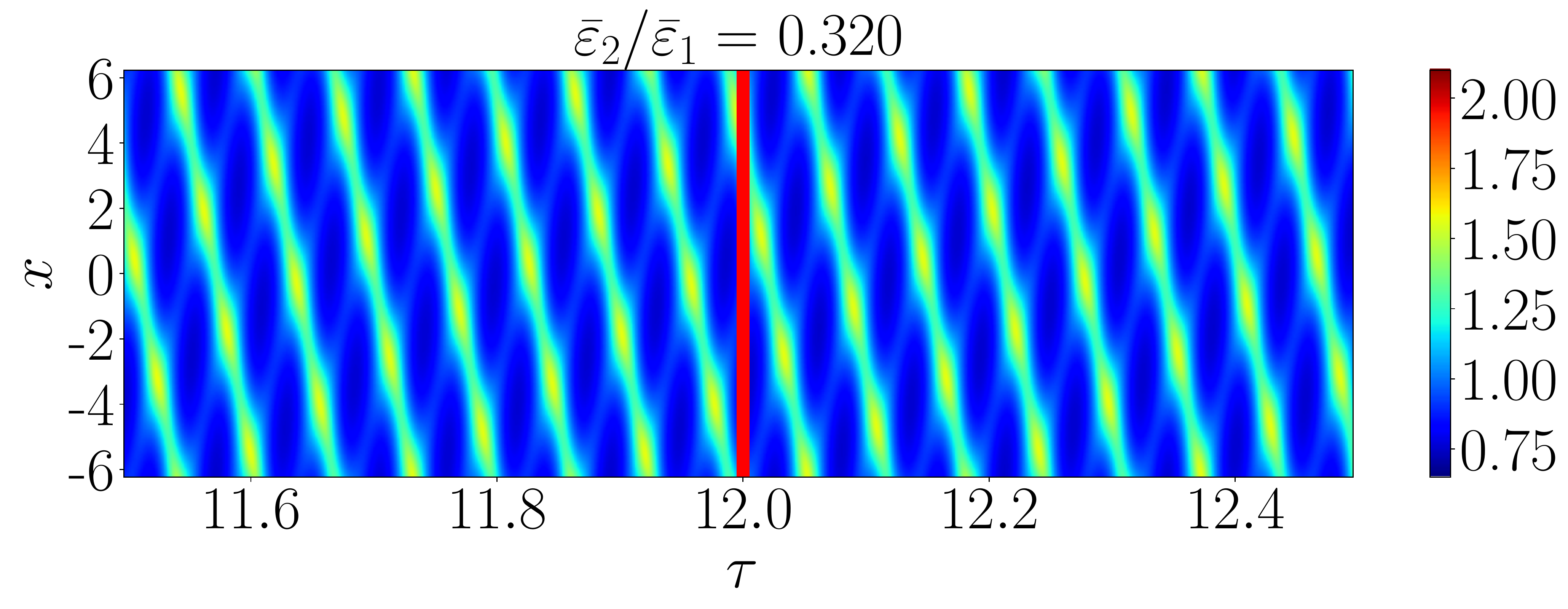}

\caption{\label{fig_full_threshold}Melting of a two-phase quasicrystal into
a single phase quasicrystal around the threshold in the fully nonlinear
model. The colormaps show the electron density $n_{e}\left(x,\tau\right)$
in the $\left(x,\tau\right)$ plane for a two-phase ion acoustic wave
excited by two driving counter propagating traveling waves with $k_{1}=-0.5$
and $k_{2}=1$ obtained by solving the fully nonlinear equations~(\ref{eq:n_t})\textendash (\ref{eq:phi_xx})
for various values of $\bar{\varepsilon}_{2}$: just above the threshold
($\bar{\varepsilon}_{2}/\bar{\varepsilon}_{1}=0.338$, top subplot),
just below the threshold ($\bar{\varepsilon}_{2}/\bar{\varepsilon}_{1}=0.337$,
middle subplot), and below the threshold ($\bar{\varepsilon}_{2}/\bar{\varepsilon}_{1}=0.32$,
bottom subplot). The parameters used in the simulations are otherwise
the same as in Figs.~\hyperref[fig_full]{\ref{fig_full}(a)},~\ref{fig_max_u},
and~\ref{fig_weakly}. The red vertical line indicates the termination
of the drive at $\tau=12$.}
\end{figure}
\begin{figure}
\includegraphics[width=1\columnwidth]{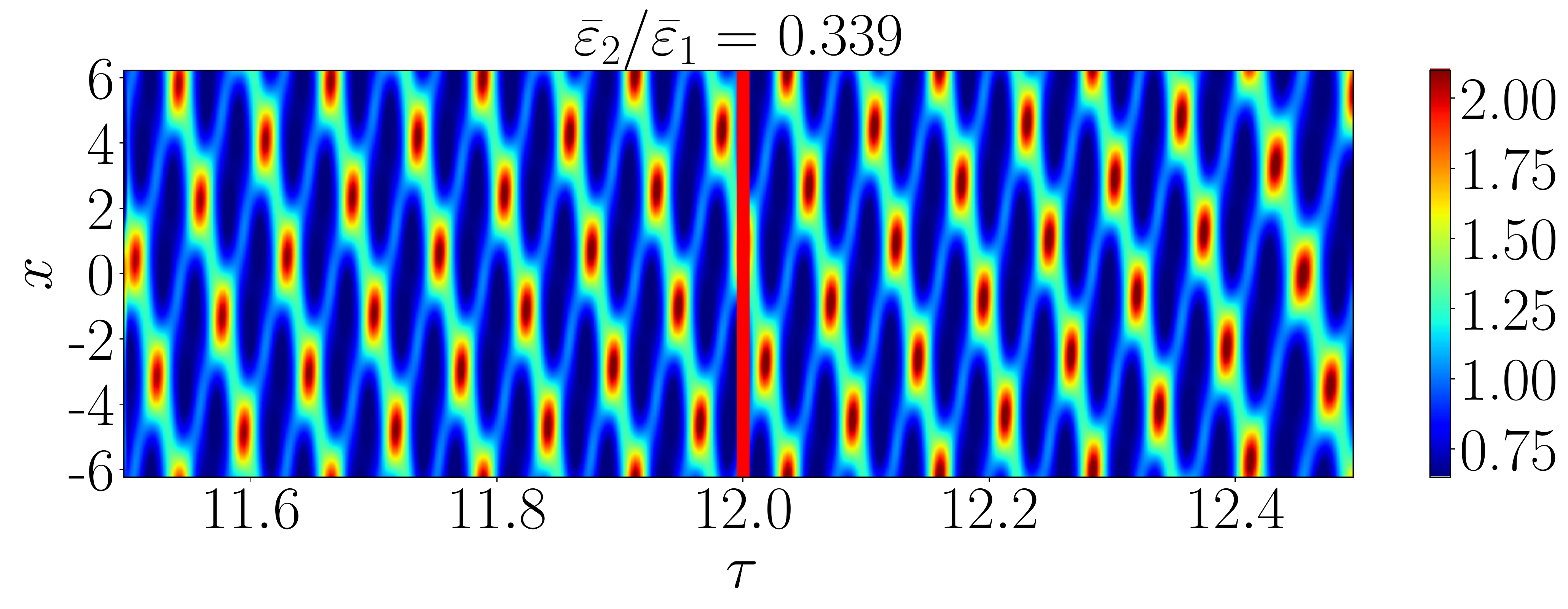}

\includegraphics[width=1\columnwidth]{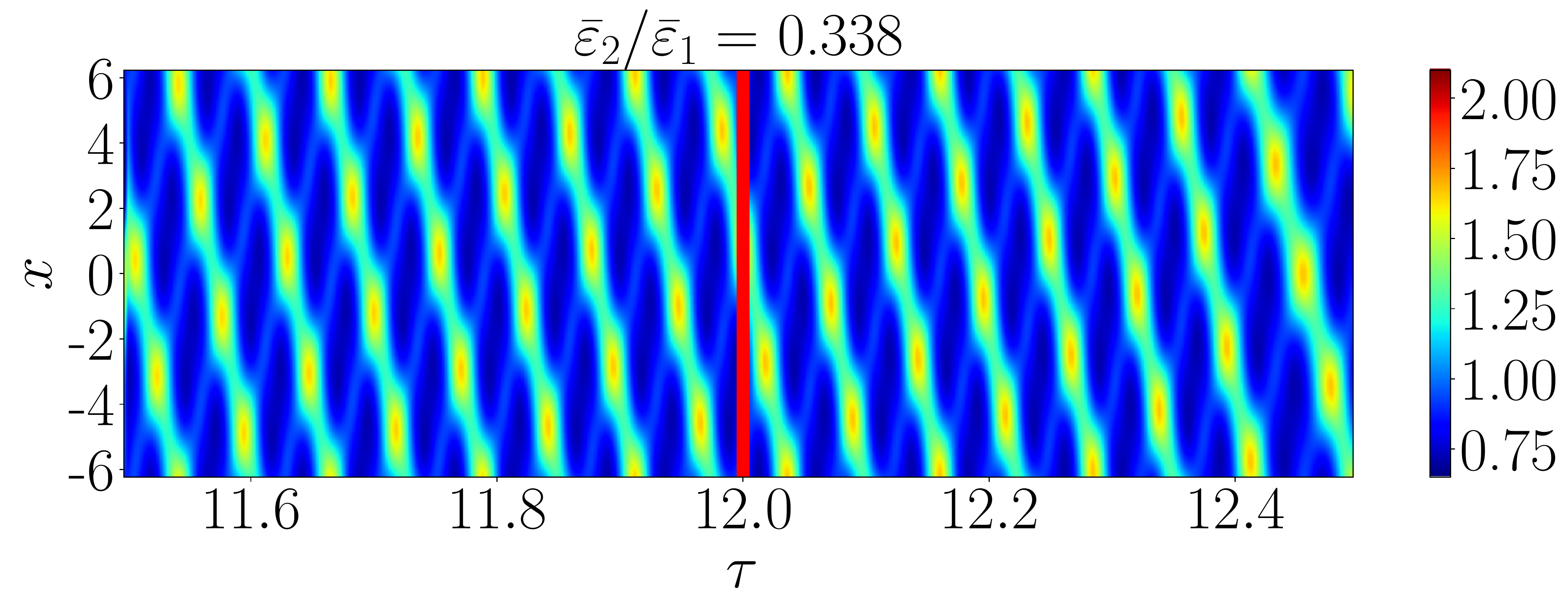}

\includegraphics[width=1\columnwidth]{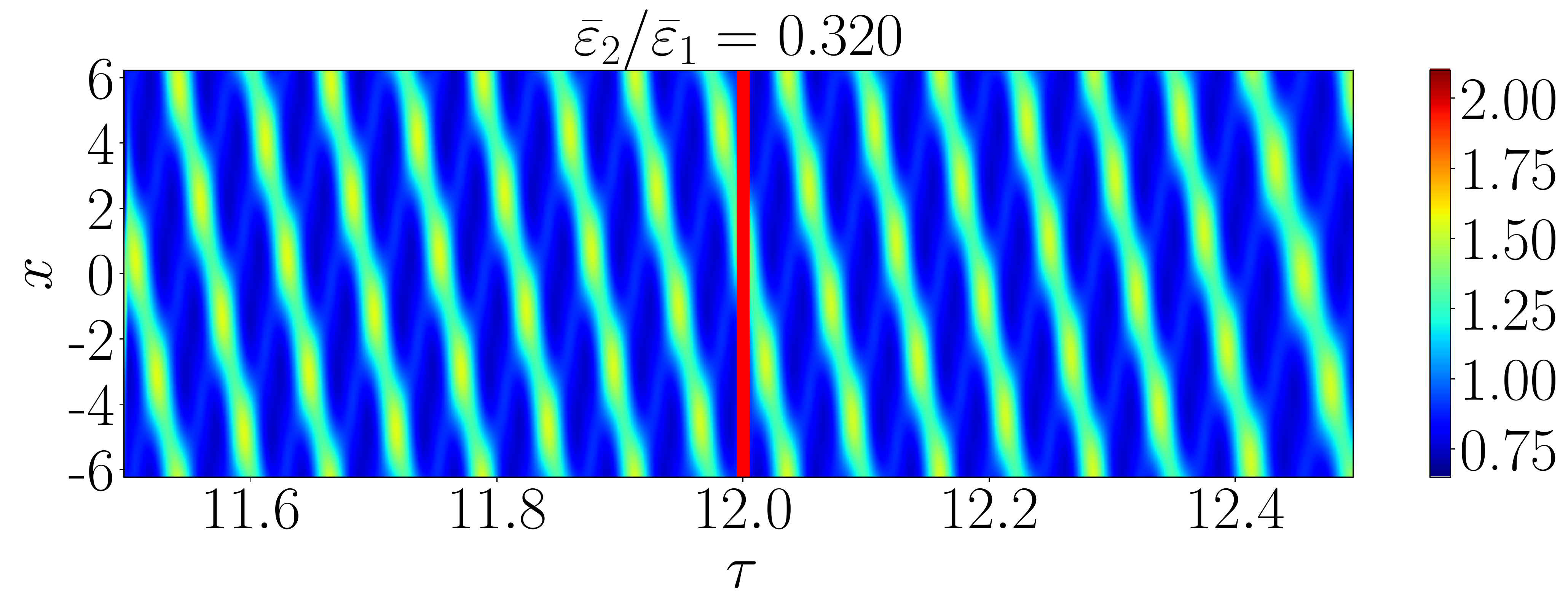}

\caption{\label{fig_weakly_threshold}Melting of a two-phase quasicrystal into
a single phase quasicrystal around the threshold in the weakly nonlinear
model. The colormaps show the electron density $n_{e}\left(x,\tau\right)$
in the $\left(x,\tau\right)$ plane for a two-phase ion acoustic wave
excited by two driving counter propagating traveling waves with $k_{1}=-0.5$
and $k_{2}=1$ obtained by solving the weakly nonlinear reduced dynamical
equations~(\ref{eq:dI1_dt})\textendash (\ref{eq:dPHI2_dt}) for
various values of $\bar{\varepsilon}_{2}$: just above the threshold
($\bar{\varepsilon}_{2}/\bar{\varepsilon}_{1}=0.339$, top subplot),
just below the threshold ($\bar{\varepsilon}_{2}/\bar{\varepsilon}_{1}=0.338$,
middle subplot), and below the threshold ($\bar{\varepsilon}_{2}/\bar{\varepsilon}_{1}=0.32$,
bottom subplot). The parameters used in the simulations are otherwise
the same as in Figs.~\hyperref[fig_full]{\ref{fig_full}(a)},~\ref{fig_max_u},
and~\ref{fig_weakly}. The red vertical line indicates the termination
of the drive at $\tau=12$.}
\end{figure}

\subsection*{Thresholds}

\noindent 
\begin{figure*}[t]
\includegraphics[width=1\columnwidth]{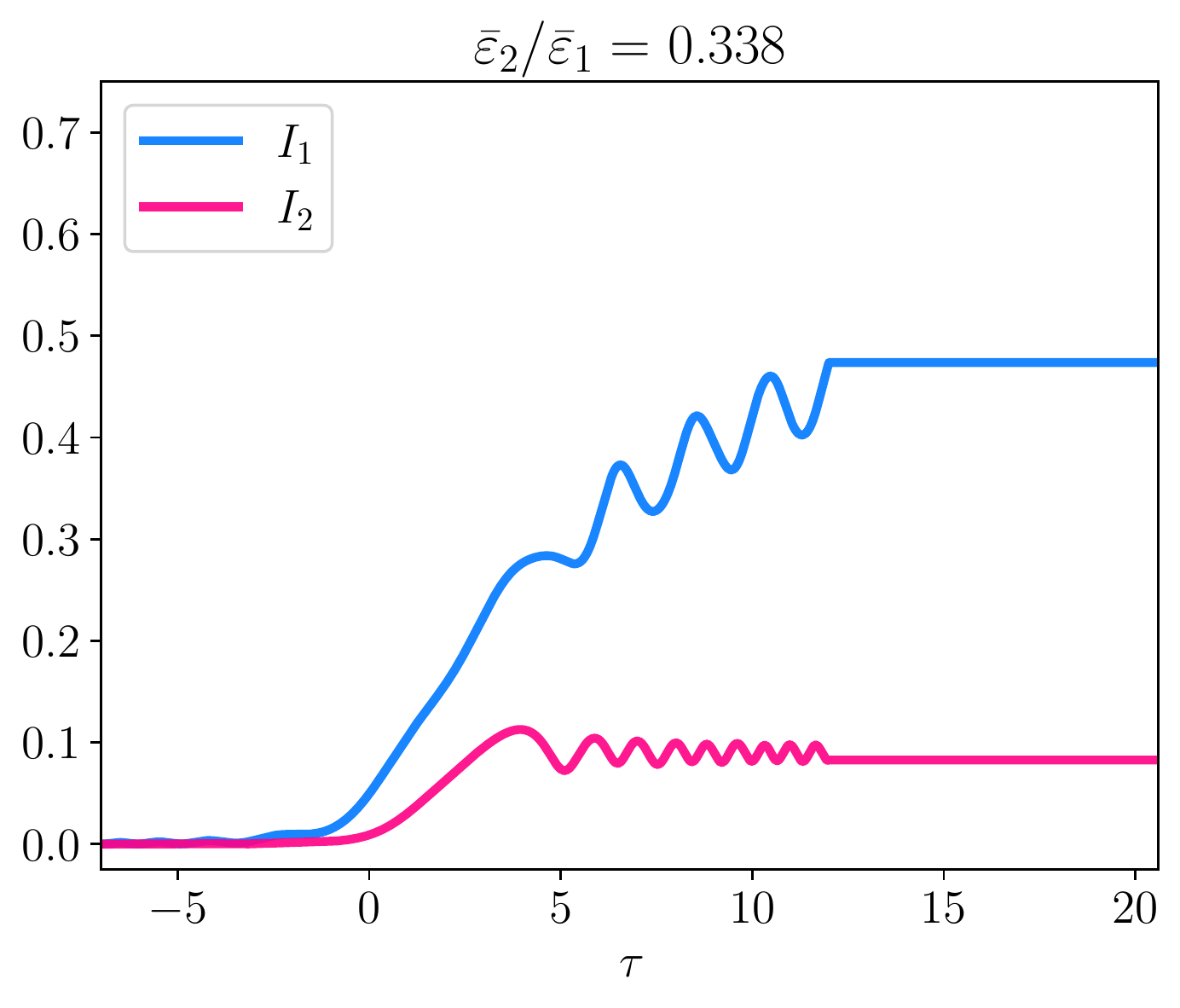}\hfill{}\includegraphics[width=1\columnwidth]{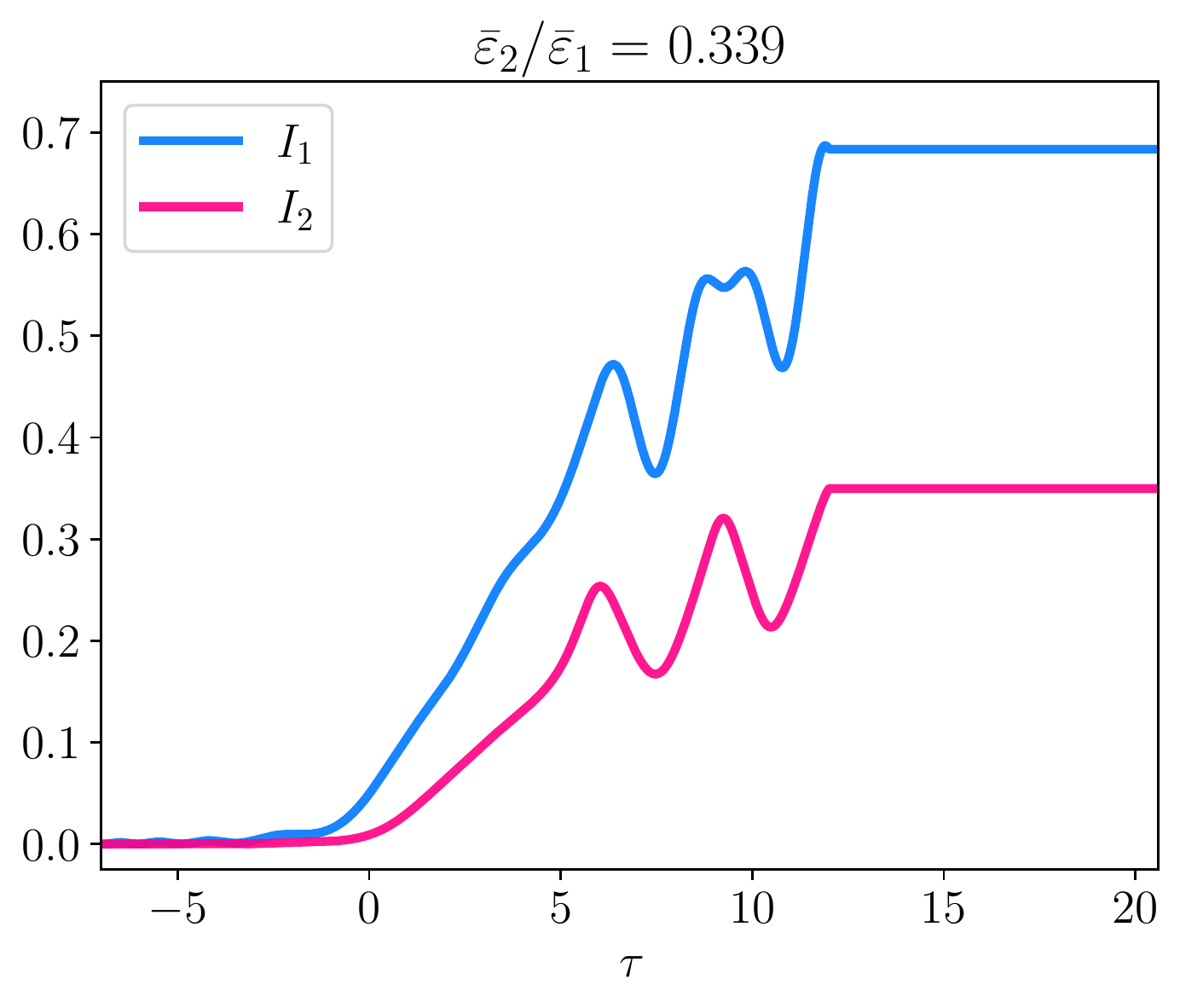}

\includegraphics[width=1\columnwidth]{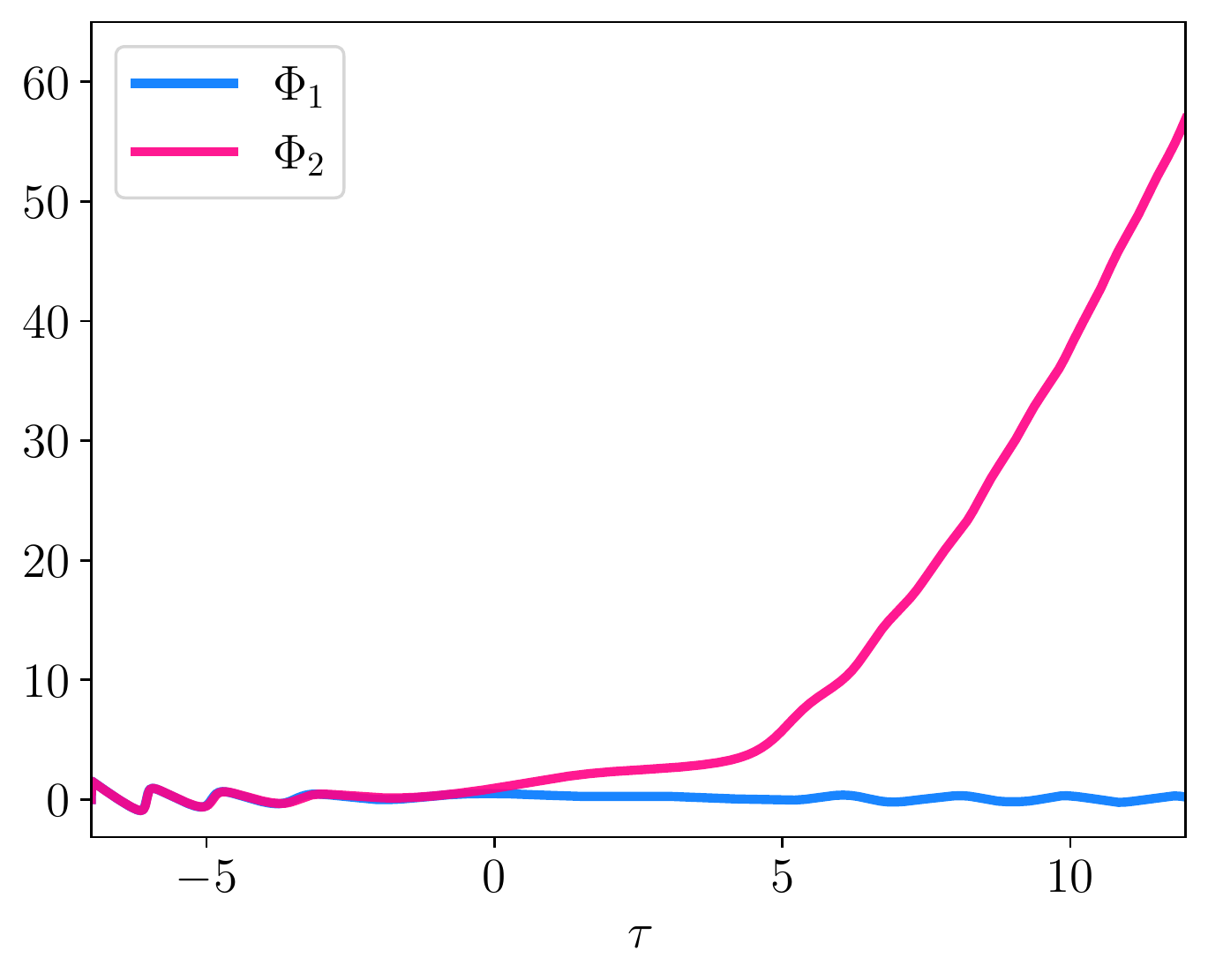}\hfill{}\includegraphics[width=1\columnwidth]{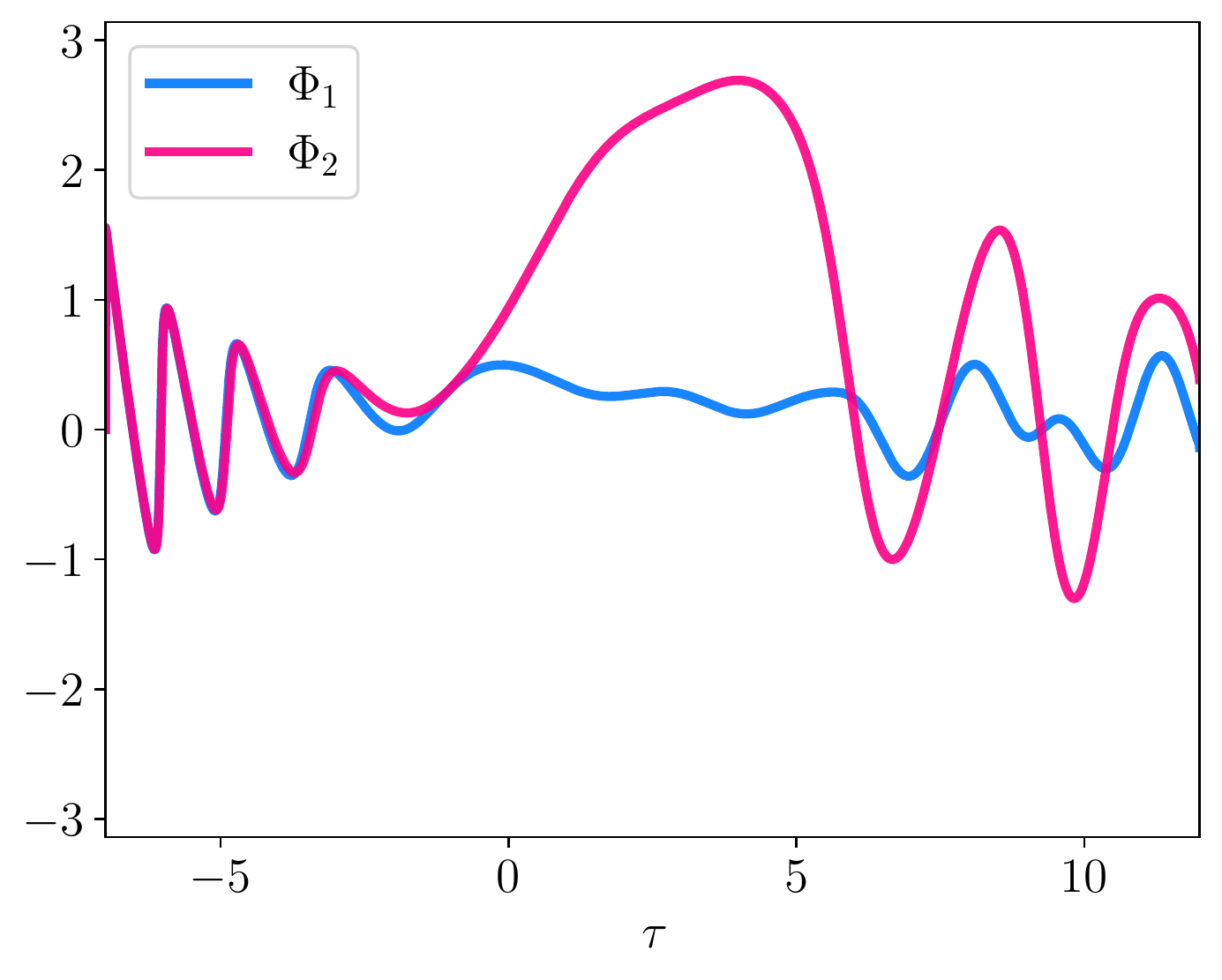}

\caption{\label{fig_I_Phi_eps_weakly}The effective actions $I_{1},I_{2}$
(top subplots) and the phase mismatches $\Phi_{1},\Phi_{2}$ (bottom
subplots) versus slow time $\tau=\sqrt{\alpha_{1}}t$ obtained by
solving the weakly nonlinear reduced dynamical equations~(\ref{eq:dI1_dt})\textendash (\ref{eq:dPHI2_dt}).
The parameters used in the simulations are the same as in Figs.~\hyperref[fig_full]{\ref{fig_full}(a)},~\ref{fig_max_u},
and~\ref{fig_weakly} except for parameter $\bar{\varepsilon}_{2}$,
which is equal to $\bar{\varepsilon}_{2}=0.338\bar{\varepsilon}_{1}$
(left side) and $\bar{\varepsilon}_{2}=0.339\bar{\varepsilon}_{1}$
(right side).}
\end{figure*}

The important dynamical characteristic of the autoresonance is that
when starting from zero, the chirped-driven system is captured into
resonance only if the driving amplitudes exceed a certain threshold
\citep{Barth2007}. Thus, if we increase the amplitudes of the drives,
the autoresonant excitation of the quasicrystals occurs abruptly,
resembling a phase transition.

The thresholds can be analyzed by reducing the problem to the motion
of pseudoparticles in an anharmonic slowly varying potential well,
similar to the way it was done in Ref.~\citep{Fajans2001} for the
single phase weakly nonlinear theory. However, unlike the threshold
for the autoresonance of a single phase ion acoustic wave (see Ref.~\citep{Friedland2014}),
the general analytical result for the double autoresonance thresholds
is difficult to obtain and the thresholds are complicated functions
of $\alpha_{1}$, $\alpha_{2}$, $a\left(k_{1}\right)$, $b\left(k_{1},k_{2}\right)$,
$c\left(k_{2}\right)$. Nonetheless, it is possible to find the thresholds
numerically. As an example let us use the same parameters as in Figs.~\hyperref[fig_full]{\ref{fig_full}(a)},~\ref{fig_max_u},~\ref{fig_weakly},
and~\ref{fig_I_Phi_eps=00003D1.0}, fix the value of $\bar{\varepsilon}_{1}$,
and solve numerically both the fully and weakly nonlinear equations
while sweeping through values of $\bar{\varepsilon}_{2}$. The results
of these simulations are presented in Figs.~\ref{fig_full_threshold}
and~\ref{fig_weakly_threshold}. Figure~\ref{fig_full_threshold}
shows a colormap of the electron density $n_{e}\left(x,\tau\right)\approx e^{\varphi\left(x,\tau\right)}$
obtained by solving the fully nonlinear equations~(\ref{eq:n_t})\textendash (\ref{eq:phi_xx})
for various values of $\bar{\varepsilon}_{2}$: just above the threshold
at $\bar{\varepsilon}_{2}=0.338\bar{\varepsilon}_{1}$ (top subplot),
just below the threshold at $\bar{\varepsilon}_{2}=0.337\bar{\varepsilon}_{1}$
(middle subplot), and below the threshold at $\bar{\varepsilon}_{2}=0.32\bar{\varepsilon}_{1}$
(bottom subplot), while Fig.~\ref{fig_weakly_threshold} shows a
colormap of the electron density $n_{e}\left(x,\tau\right)\approx e^{\varphi\left(x,\tau\right)}$
obtained by solving the weakly nonlinear equations~(\ref{eq:dI1_dt})\textendash (\ref{eq:dPHI2_dt})
for various values of $\bar{\varepsilon}_{2}$: just above the threshold
($\bar{\varepsilon}_{2}/\bar{\varepsilon}_{1}=0.339$, top subplot),
just below the threshold ($\bar{\varepsilon}_{2}/\bar{\varepsilon}_{1}=0.338$,
middle subplot), and below the threshold ($\bar{\varepsilon}_{2}/\bar{\varepsilon}_{1}=0.32$,
bottom subplot). We can see from these figures that the crystallization
is indeed an abrupt phenomenon akin to a phase transition. For the
fully nonlinear simulations the threshold lies between $\bar{\varepsilon}_{2}/\bar{\varepsilon}_{1}=0.337$
and $\bar{\varepsilon}_{2}/\bar{\varepsilon}_{1}=0.338$, while for
the weakly nonlinear simulations the threshold lies between $\bar{\varepsilon}_{2}/\bar{\varepsilon}_{1}=0.338$
and $\bar{\varepsilon}_{2}/\bar{\varepsilon}_{1}=0.339$. Such a good
agreement for the value of the threshold is another proof that our
weakly nonlinear theory is applicable. To better illustrate the threshold
nature of the double autoresonance we also produced an animation showing
the crystallization of a single phase structure into a two-phase structure
around the threshold as we sweep $\bar{\varepsilon}_{2}$ from $\bar{\varepsilon}_{2}=0.32\bar{\varepsilon}_{1}$
to $\bar{\varepsilon}_{2}=0.35\bar{\varepsilon}_{1}$. The animation
is available in the Supplemental Material \citep{animation}.

The necessity of the phase locking for the autoresonance is demonstrated
in Fig.~\ref{fig_I_Phi_eps_weakly}, which shows the effective actions
$I_{1},I_{2}$ and the phase mismatches $\Phi_{1},\Phi_{2}$ as functions
of slow time $\tau=\sqrt{\alpha_{1}}t$ just below the threshold ($\bar{\varepsilon}_{2}/\bar{\varepsilon}_{1}=0.338$,
left side) and just above the threshold ($\bar{\varepsilon}_{2}/\bar{\varepsilon}_{1}=0.339$,
right side). One can clearly see that just below the threshold the
second action $I_{2}$ does not enter the autoresonant regime and
the growth of the amplitude saturates, while the phase mismatch $\Phi_{2}$
increases with time signifying the absence of phase locking. In contrast,
just above the threshold both phases are locked and the amplitudes
$I_{1},I_{2}$ increase in time similar to the case shown in Fig.~\ref{fig_I_Phi_eps=00003D1.0}.

\section{\label{sec_Conclusion}Conclusions}

We have demonstrated by means of nonlinear numerical simulations that
it is possible to create quasicrystalline spatiotemporal structures
in plasma by exciting large amplitude two-phase ion acoustic waves
nonlinearly phased locked into the corresponding small amplitude traveling
wave drives with chirped frequencies. We have used the Lagrangian
formulation and Whitham's averaged variational method to derive analytical
results describing the weakly nonlinear evolution of the system. We
have applied the weakly nonlinear analytical theory to determine the
parameters necessary for the successful excitation and control of
multiphase waves. The nonlinearly excited quasicrystalline structures
remain even after we turn off the drive. Thus, the space-time quasicrystalline
structure in plasma can be excited and then used independently for
the purpose of plasma photonics experiments. While our warm fluid
model does not have dissipation or noise due to collisions or Landau
damping, for example, and, generally speaking, its long-time stability
must be addressed by kinetic or particle-in-cell (PIC) simulations,
it is apparent that, depending on the time scales of the problem,
this structure can be considered as at least a dissipative space-time
crystal. We also note that we do not address in this paper whether
the driven space-time crystal is a ``true'' time crystal in the
sense of spontaneous symmetry breaking as proposed in Refs.~\citep{Wilczek2012,Shapere2012};
for more discussion regarding this see Refs.~\citep{Yao2020,Sacha2020,Else2020}.

It is expected that, using our technique, similar structures can be
driven in other systems, for example dust acoustic waves in complex
plasmas \citep{Rao1990}. We expect that multiphase solutions when
the number of drives exceeds two are also possible; however, it will
be harder to analyze such a system analytically. Finally, we point
out that beyond any practical application as a plasma photonic (or
accelerating) structure, the possibility of exciting multiphase solutions
for in general non-integrable warm ion acoustic waves system is in
itself an important fundamental result in the field of nonlinear dynamics.
These techniques developed for the excitation of large amplitude ion
acoustic waves may be applied to other partial differential equation
systems that can support multiphase nonlinear waves.

\section*{Acknowledgments}

This work was supported by NSF-BSF Grant No. 1803874 and US-Israel
Binational Science Foundation Grant No. 2020233.

\begin{widetext}

\appendix

\section{The averaged Lagrangian density} 
\label{Appendix_A}

The averaged Lagrangian density
\begin{equation}
\bar{L}=\left\langle L\left(\theta_{1},\theta_{2},t\right)\right\rangle _{\theta_{1},\theta_{2}}=\int L\left(\theta_{1},\theta_{2},t\right)\frac{d\theta_{1}}{2\pi}\frac{d\theta_{2}}{2\pi}
\end{equation}
is equal to the sum of the following terms:

\begin{equation}
\left\langle \frac{1}{2}\varphi_{x}^{2}\right\rangle _{\theta_{1},\theta_{2}}=\frac{1}{4}k_{1}^{2}C_{10}^{2}+\frac{1}{4}k_{2}^{2}C_{01}^{2}+\frac{1}{4}\left(k_{1}+k_{2}\right)^{2}C_{11}^{2}+\frac{1}{4}\left(k_{1}-k_{2}\right)^{2}C_{1,-1}^{2}+k_{1}^{2}C_{20}^{2}+k_{2}^{2}C_{02}^{2},
\end{equation}

\begin{equation}
\left\langle \varphi\right\rangle _{\theta_{1},\theta_{2}}=C_{00},
\end{equation}

\begin{equation}
\left\langle \frac{1}{2}\varphi^{2}\right\rangle _{\theta_{1},\theta_{2}}=\frac{1}{2}C_{00}^{2}+\frac{1}{4}C_{10}^{2}+\frac{1}{4}C_{01}^{2}+\frac{1}{4}C_{11}^{2}+\frac{1}{4}C_{1,-1}^{2}+\frac{1}{4}C_{20}^{2}+\frac{1}{4}C_{02}^{2},
\end{equation}

\begin{equation}
\left\langle \frac{1}{6}\varphi^{3}\right\rangle _{\theta_{1},\theta_{2}}=\frac{1}{4}\left(C_{00}+\frac{1}{2}C_{02}\right)C_{01}^{2}+\frac{1}{4}\left(C_{00}+\frac{1}{2}C_{20}\right)C_{10}^{2}+\frac{1}{4}\left(C_{1,-1}+C_{11}\right)C_{01}C_{10},
\end{equation}

\begin{equation}
\left\langle \frac{1}{24}\varphi^{4}\right\rangle _{\theta_{1},\theta_{2}}=\frac{1}{64}C_{01}^{4}+\frac{1}{16}C_{01}^{2}C_{10}^{2}+\frac{1}{64}C_{10}^{4},
\end{equation}

\begin{multline}
\left\langle -\frac{1}{2}\left(\psi_{t}\sigma_{x}+\psi_{x}\sigma_{t}\right)\right\rangle _{\theta_{1},\theta_{2}}=\frac{1}{2}\omega_{1}k_{1}\tilde{B}_{10}\tilde{A}_{10}+\frac{1}{2}\omega_{2}k_{2}\tilde{B}_{01}\tilde{A}_{01}\\
+2\omega_{1}k_{1}\tilde{B}_{20}\tilde{A}_{20}+2\omega_{2}k_{2}\tilde{B}_{02}\tilde{A}_{02}\\
+\frac{1}{2}\left(\omega_{1}-\omega_{2}\right)\left(k_{1}-k_{2}\right)\tilde{B}_{1,-1}\tilde{A}_{1,-1}+\frac{1}{2}\left(\omega_{1}+\omega_{2}\right)\left(k_{1}+k_{2}\right)\tilde{B}_{11}\tilde{A}_{11},
\end{multline}

\begin{multline}
\left\langle -\left(\frac{1}{2}\psi_{x}^{2}+\varphi\right)\left(1+\sigma_{x}\right)\right\rangle _{\theta_{1},\theta_{2}}=-C_{00}-\frac{1}{4}k_{1}^{2}\tilde{B}_{10}^{2}-\frac{1}{4}k_{2}^{2}\tilde{B}_{01}^{2}-k_{1}^{2}\tilde{B}_{20}^{2}-k_{2}^{2}\tilde{B}_{02}^{2}-\frac{1}{4}\left(k_{1}+k_{2}\right)^{2}\tilde{B}_{11}^{2}-\frac{1}{4}\left(k_{1}-k_{2}\right)^{2}\tilde{B}_{1,-1}^{2}\\
-k_{1}\left(\frac{1}{2}C_{10}\tilde{A}_{10}+C_{20}\tilde{A}_{20}\right)-k_{2}\left(\frac{1}{2}C_{01}\tilde{A}_{01}+C_{02}\tilde{A}_{02}\right)-\frac{1}{2}\left(k_{1}+k_{2}\right)C_{11}\tilde{A}_{11}-\frac{1}{2}\left(k_{1}-k_{2}\right)C_{1,-1}\tilde{A}_{1,-1}\\
-\frac{1}{2}k_{2}^{3}\left(\tilde{A}_{01}\tilde{B}_{01}\tilde{B}_{02}+\frac{1}{2}\tilde{A}_{02}\tilde{B}_{01}^{2}\right)-\frac{1}{2}k_{1}^{3}\left(\tilde{A}_{10}\tilde{B}_{10}\tilde{B}_{20}+\frac{1}{2}\tilde{A}_{20}\tilde{B}_{10}^{2}\right)\\
-\frac{1}{4}k_{1}k_{2}\left(k_{1}-k_{2}\right)\left(\tilde{A}_{01}\tilde{B}_{10}\tilde{B}_{1,-1}+\tilde{A}_{1,-1}\tilde{B}_{01}\tilde{B}_{10}+\tilde{A}_{10}\tilde{B}_{01}\tilde{B}_{1,-1}\right)\\
-\frac{1}{4}k_{1}k_{2}\left(k_{1}+k_{2}\right)\left(\tilde{A}_{01}\tilde{B}_{11}\tilde{B}_{10}+\tilde{A}_{10}\tilde{B}_{01}\tilde{B}_{11}+\tilde{A}_{11}\tilde{B}_{01}\tilde{B}_{10}\right),
\end{multline}

\begin{multline}
\left\langle -\frac{\Delta^{2}}{2}\sigma_{x}^{2}\left(1+\frac{1}{3}\sigma_{x}\right)\right\rangle _{\theta_{1},\theta_{2}}=-\frac{\Delta^{2}}{4}k_{1}^{2}\tilde{A}_{10}^{2}-\frac{\Delta^{2}}{4}k_{1}^{3}\tilde{A}_{10}^{2}\tilde{A}_{20}-\Delta^{2}k_{1}^{2}\tilde{A}_{20}^{2}-\frac{\Delta^{2}}{4}k_{2}^{2}\tilde{A}_{01}^{2}-\frac{\Delta^{2}}{4}k_{2}^{3}\tilde{A}_{01}^{2}\tilde{A}_{02}-\Delta^{2}k_{2}^{2}\tilde{A}_{02}^{2}\\
-\frac{\Delta^{2}}{4}k_{1}k_{2}\left(k_{1}-k_{2}\right)\tilde{A}_{01}\tilde{A}_{10}\tilde{A}_{1,-1}-\frac{\Delta^{2}}{4}k_{1}k_{2}\left(k_{1}+k_{2}\right)\tilde{A}_{01}\tilde{A}_{10}\tilde{A}_{11}-\frac{\Delta^{2}}{4}\left(k_{1}-k_{2}\right)^{2}\tilde{A}_{1,-1}^{2}-\frac{\Delta^{2}}{4}\left(k_{1}+k_{2}\right)^{2}\tilde{A}_{11}^{2},
\end{multline}

\begin{equation}
\left\langle \varphi\varphi_{d}\right\rangle _{\theta_{1},\theta_{2}}=\frac{\varepsilon_{1}}{2}C_{10}\cos\left(\Phi_{1}\right)+\frac{\varepsilon_{2}}{2}C_{01}\cos\left(\Phi_{2}\right).
\end{equation}

\section{The variations and the amplitudes} 
\label{Appendix_B}

In this Appendix we calculate the variations of the averaged Lagrangian
with respect to the various amplitudes and express all the second-order
amplitudes $\tilde{A}_{20}$, $\tilde{A}_{02}$, $\tilde{B}_{20}$,
$\tilde{B}_{02}$, $C_{20}$, $C_{02}$, $\tilde{A}_{11}$, $\tilde{A}_{1,-1}$,
$\tilde{B}_{11}$, $\tilde{B}_{1,-1}$, $C_{11}$, $C_{1,-1}$ through
the first-order amplitudes $C_{10}$ and $C_{01}$.

To express the second-order amplitudes through $C_{10}$ and $C_{01}$
let us first calculate the variations of the averaged Lagrangian density
with respect to the second-order amplitudes:

\begin{equation}
\frac{\partial\bar{L}}{\partial C_{00}}=C_{00}+\frac{1}{4}C_{01}^{2}+\frac{1}{4}C_{10}^{2}=0,\label{eq:C00}
\end{equation}

\begin{equation}
\frac{\partial\bar{L}}{\partial C_{20}}=2k_{1}^{2}C_{20}+\frac{1}{2}C_{20}+\frac{1}{8}C_{10}^{2}-k_{1}\tilde{A}_{20}=0,\label{eq:C20}
\end{equation}

\begin{equation}
\frac{\partial\bar{L}}{\partial C_{02}}=2k_{2}^{2}C_{02}+\frac{1}{2}C_{02}+\frac{1}{8}C_{01}^{2}-k_{2}\tilde{A}_{02}=0,\label{eq:C02}
\end{equation}

\begin{equation}
\frac{\partial\bar{L}}{\partial C_{11}}=\frac{1}{2}\left(k_{1}+k_{2}\right)^{2}C_{11}+\frac{1}{2}C_{11}+\frac{1}{4}C_{01}C_{10}-\frac{1}{2}\left(k_{1}+k_{2}\right)\tilde{A}_{11}=0,\label{eq:C11}
\end{equation}

\begin{equation}
\frac{\partial\bar{L}}{\partial C_{1,-1}}=\frac{1}{2}\left(k_{1}-k_{2}\right)^{2}C_{1,-1}+\frac{1}{2}C_{1,-1}+\frac{1}{4}C_{01}C_{10}-\frac{1}{2}\left(k_{1}-k_{2}\right)\tilde{A}_{1,-1}=0,\label{eq:C1-1}
\end{equation}

\begin{equation}
\frac{\partial\bar{L}}{\partial\tilde{A}_{20}}=2\omega_{1}k_{1}\tilde{B}_{20}-k_{1}C_{20}-\frac{1}{4}k_{1}^{3}\tilde{B}_{10}^{2}-\frac{\Delta^{2}}{4}k_{1}^{3}\tilde{A}_{10}^{2}-2\Delta^{2}k_{1}^{2}\tilde{A}_{20}=0,\label{eq:A20}
\end{equation}

\begin{equation}
\frac{\partial\bar{L}}{\partial\tilde{A}_{02}}=2\omega_{2}k_{2}\tilde{B}_{02}-k_{2}C_{02}-\frac{1}{4}k_{2}^{3}\tilde{B}_{01}^{2}-\frac{\Delta^{2}}{4}k_{2}^{3}\tilde{A}_{01}^{2}-2\Delta^{2}k_{2}^{2}\tilde{A}_{02}=0,\label{eq:A02}
\end{equation}

\begin{multline}
\frac{\partial\bar{L}}{\partial\tilde{A}_{11}}=\frac{1}{2}\left(\omega_{1}+\omega_{2}\right)\left(k_{1}+k_{2}\right)\tilde{B}_{11}-\frac{1}{2}\left(k_{1}+k_{2}\right)C_{11}-\frac{1}{4}k_{1}k_{2}\left(k_{1}+k_{2}\right)\tilde{B}_{01}\tilde{B}_{10}\\
-\frac{\Delta^{2}}{4}k_{1}k_{2}\left(k_{1}+k_{2}\right)\tilde{A}_{01}\tilde{A}_{10}-\frac{\Delta^{2}}{2}\left(k_{1}+k_{2}\right)^{2}\tilde{A}_{11}=0,\label{eq:A11}
\end{multline}

\begin{multline}
\frac{\partial\bar{L}}{\partial\tilde{A}_{1,-1}}=\frac{1}{2}\left(\omega_{1}-\omega_{2}\right)\left(k_{1}-k_{2}\right)\tilde{B}_{1,-1}-\frac{1}{2}\left(k_{1}-k_{2}\right)C_{1,-1}-\frac{1}{4}k_{1}k_{2}\left(k_{1}-k_{2}\right)\tilde{B}_{01}\tilde{B}_{10}\\
-\frac{\Delta^{2}}{4}k_{1}k_{2}\left(k_{1}-k_{2}\right)\tilde{A}_{01}\tilde{A}_{10}-\frac{\Delta^{2}}{2}\left(k_{1}-k_{2}\right)^{2}\tilde{A}_{1,-1}=0,\label{eq:A1-1}
\end{multline}

\begin{equation}
\frac{\partial\bar{L}}{\partial\tilde{B}_{20}}=2\omega_{1}k_{1}\tilde{A}_{20}-2k_{1}^{2}\tilde{B}_{20}-\frac{1}{2}k_{1}^{3}\tilde{A}_{10}\tilde{B}_{10}=0,\label{eq:B20}
\end{equation}

\begin{equation}
\frac{\partial\bar{L}}{\partial\tilde{B}_{02}}=2\omega_{2}k_{2}\tilde{A}_{02}-2k_{2}^{2}\tilde{B}_{02}-\frac{1}{2}k_{2}^{3}\tilde{A}_{01}\tilde{B}_{01}=0,\label{eq:B02}
\end{equation}

\begin{equation}
\frac{\partial\bar{L}}{\partial\tilde{B}_{11}}=\frac{1}{2}\left(\omega_{1}+\omega_{2}\right)\left(k_{1}+k_{2}\right)\tilde{A}_{11}-\frac{1}{2}\left(k_{1}+k_{2}\right)^{2}\tilde{B}_{11}-\frac{1}{4}k_{1}k_{2}\left(k_{1}+k_{2}\right)\left(\tilde{A}_{01}\tilde{B}_{10}+\tilde{A}_{10}\tilde{B}_{01}\right)=0,\label{eq:B11}
\end{equation}

\begin{equation}
\frac{\partial\bar{L}}{\partial\tilde{B}_{1,-1}}=\frac{1}{2}\left(\omega_{1}-\omega_{2}\right)\left(k_{1}-k_{2}\right)\tilde{A}_{1,-1}-\frac{1}{2}\left(k_{1}-k_{2}\right)^{2}\tilde{B}_{1,-1}-\frac{1}{4}k_{1}k_{2}\left(k_{1}-k_{2}\right)\left(\tilde{A}_{01}\tilde{B}_{10}+\tilde{A}_{10}\tilde{B}_{01}\right)=0.\label{eq:B1-1}
\end{equation}

From Eq.~(\ref{eq:C00}) we obtain

\begin{equation}
C_{00}=-\frac{1}{4}C_{01}^{2}-\frac{1}{4}C_{10}^{2}.\label{eq:C00_2}
\end{equation}

From Eqs.~(\ref{eq:C20}), (\ref{eq:C02}), (\ref{eq:A20}), (\ref{eq:A02}),
(\ref{eq:B20}), and~(\ref{eq:B02}) we obtain

\begin{equation}
\tilde{A}_{20}=\frac{k_{1}^{3}\left(4k_{1}^{2}+1\right)\tilde{B}_{10}^{2}+2\omega_{1}k_{1}^{2}\left(4k_{1}^{2}+1\right)\tilde{A}_{10}\tilde{B}_{10}-k_{1}C_{10}^{2}+\Delta^{2}k_{1}^{3}\left(4k_{1}^{2}+1\right)\tilde{A}_{10}^{2}}{8\left[\left(\omega_{1}^{2}-\Delta^{2}k_{1}^{2}\right)\left(4k_{1}^{2}+1\right)-k_{1}^{2}\right]},
\end{equation}

\begin{equation}
\tilde{A}_{02}=\frac{k_{2}^{3}\left(4k_{2}^{2}+1\right)\tilde{B}_{01}^{2}+2\omega_{2}k_{2}^{2}\left(4k_{2}^{2}+1\right)\tilde{A}_{01}\tilde{B}_{01}-k_{2}C_{01}^{2}+\Delta^{2}k_{2}^{3}\left(4k_{2}^{2}+1\right)\tilde{A}_{01}^{2}}{8\left[\left(\omega_{2}^{2}-\Delta^{2}k_{2}^{2}\right)\left(4k_{2}^{2}+1\right)-k_{2}^{2}\right]},
\end{equation}

\begin{equation}
\tilde{B}_{20}=\frac{\omega_{1}k_{1}^{2}\left(4k_{1}^{2}+1\right)\tilde{B}_{10}^{2}+2k_{1}^{3}\tilde{A}_{10}\tilde{B}_{10}-\omega_{1}C_{10}^{2}+\Delta^{2}\omega_{1}k_{1}^{2}\left(4k_{1}^{2}+1\right)\tilde{A}_{10}^{2}+2\Delta^{2}k_{1}^{3}\left(4k_{1}^{2}+1\right)\tilde{A}_{10}\tilde{B}_{10}}{8\left[\left(\omega_{1}^{2}-\Delta^{2}k_{1}^{2}\right)\left(4k_{1}^{2}+1\right)-k_{1}^{2}\right]},
\end{equation}

\begin{equation}
\tilde{B}_{02}=\frac{\omega_{2}k_{2}^{2}\left(4k_{2}^{2}+1\right)\tilde{B}_{01}^{2}+2k_{2}^{3}\tilde{A}_{01}\tilde{B}_{01}-\omega_{2}C_{01}^{2}+\Delta^{2}\omega_{2}k_{2}^{2}\left(4k_{2}^{2}+1\right)\tilde{A}_{01}^{2}+2\Delta^{2}k_{2}^{3}\left(4k_{2}^{2}+1\right)\tilde{A}_{01}\tilde{B}_{01}}{8\left[\left(\omega_{2}^{2}-\Delta^{2}k_{2}^{2}\right)\left(4k_{2}^{2}+1\right)-k_{2}^{2}\right]},
\end{equation}

\begin{equation}
C_{20}=\frac{k_{1}^{4}\tilde{B}_{10}^{2}+2\omega_{1}k_{1}^{3}\tilde{A}_{10}\tilde{B}_{10}-\omega_{1}^{2}C_{10}^{2}+\Delta^{2}k_{1}^{4}\tilde{A}_{10}^{2}+\Delta^{2}k_{1}^{2}C_{10}^{2}}{4\left[\left(\omega_{1}^{2}-\Delta^{2}k_{1}^{2}\right)\left(4k_{1}^{2}+1\right)-k_{1}^{2}\right]},
\end{equation}

\begin{equation}
C_{02}=\frac{k_{2}^{4}\tilde{B}_{01}^{2}+2\omega_{2}k_{2}^{3}\tilde{A}_{01}\tilde{B}_{01}-\omega_{2}^{2}C_{01}^{2}+\Delta^{2}k_{2}^{4}\tilde{A}_{01}^{2}+\Delta^{2}k_{2}^{2}C_{01}^{2}}{4\left[\left(\omega_{2}^{2}-\Delta^{2}k_{2}^{2}\right)\left(4k_{2}^{2}+1\right)-k_{2}^{2}\right]}.
\end{equation}

From Eqs.~(\ref{eq:C11}), (\ref{eq:C1-1}), (\ref{eq:A11}), (\ref{eq:A1-1}),
(\ref{eq:B11}), and~(\ref{eq:B1-1}) we obtain

\begin{multline}
\tilde{A}_{11}=\frac{k_{1}k_{2}\left(k_{1}+k_{2}\right)\left[\left(k_{1}+k_{2}\right)^{2}+1\right]\tilde{B}_{01}\tilde{B}_{10}-\left(k_{1}+k_{2}\right)C_{01}C_{10}}{2\left\{ \left[\left(\omega_{1}+\omega_{2}\right)^{2}-\Delta^{2}\left(k_{1}+k_{2}\right)^{2}\right]\left[\left(k_{1}+k_{2}\right)^{2}+1\right]-\left(k_{1}+k_{2}\right)^{2}\right\} }\\
+\frac{\left(\omega_{1}+\omega_{2}\right)k_{1}k_{2}\left[\left(k_{1}+k_{2}\right)^{2}+1\right]\left(\tilde{A}_{01}\tilde{B}_{10}+\tilde{A}_{10}\tilde{B}_{01}\right)}{2\left\{ \left[\left(\omega_{1}+\omega_{2}\right)^{2}-\Delta^{2}\left(k_{1}+k_{2}\right)^{2}\right]\left[\left(k_{1}+k_{2}\right)^{2}+1\right]-\left(k_{1}+k_{2}\right)^{2}\right\} }\\
+\frac{\Delta^{2}k_{1}k_{2}\left(k_{1}+k_{2}\right)\left[\left(k_{1}+k_{2}\right)^{2}+1\right]\tilde{A}_{01}\tilde{A}_{10}}{2\left\{ \left[\left(\omega_{1}+\omega_{2}\right)^{2}-\Delta^{2}\left(k_{1}+k_{2}\right)^{2}\right]\left[\left(k_{1}+k_{2}\right)^{2}+1\right]-\left(k_{1}+k_{2}\right)^{2}\right\} },
\end{multline}

\begin{multline}
\tilde{A}_{1,-1}=\frac{k_{1}k_{2}\left(k_{1}-k_{2}\right)\left[\left(k_{1}-k_{2}\right)^{2}+1\right]\tilde{B}_{01}\tilde{B}_{10}-\left(k_{1}-k_{2}\right)C_{01}C_{10}}{2\left\{ \left[\left(\omega_{1}-\omega_{2}\right)^{2}-\Delta^{2}\left(k_{1}-k_{2}\right)^{2}\right]\left[\left(k_{1}-k_{2}\right)^{2}+1\right]-\left(k_{1}-k_{2}\right)^{2}\right\} }\\
+\frac{\left(\omega_{1}-\omega_{2}\right)k_{1}k_{2}\left[\left(k_{1}-k_{2}\right)^{2}+1\right]\left(\tilde{A}_{01}\tilde{B}_{10}+\tilde{A}_{10}\tilde{B}_{01}\right)}{2\left\{ \left[\left(\omega_{1}-\omega_{2}\right)^{2}-\Delta^{2}\left(k_{1}-k_{2}\right)^{2}\right]\left[\left(k_{1}-k_{2}\right)^{2}+1\right]-\left(k_{1}-k_{2}\right)^{2}\right\} }\\
+\frac{\Delta^{2}k_{1}k_{2}\left(k_{1}-k_{2}\right)\left[\left(k_{1}-k_{2}\right)^{2}+1\right]\tilde{A}_{01}\tilde{A}_{10}}{2\left\{ \left[\left(\omega_{1}-\omega_{2}\right)^{2}-\Delta^{2}\left(k_{1}-k_{2}\right)^{2}\right]\left[\left(k_{1}-k_{2}\right)^{2}+1\right]-\left(k_{1}-k_{2}\right)^{2}\right\} },
\end{multline}

\begin{multline}
\tilde{B}_{11}=\frac{\left(\omega_{1}+\omega_{2}\right)k_{1}k_{2}\left[\left(k_{1}+k_{2}\right)^{2}+1\right]\tilde{B}_{01}\tilde{B}_{10}-\left(\omega_{1}+\omega_{2}\right)C_{01}C_{10}+k_{1}k_{2}\left(k_{1}+k_{2}\right)\left(\tilde{A}_{01}\tilde{B}_{10}+\tilde{A}_{10}\tilde{B}_{01}\right)}{2\left\{ \left[\left(\omega_{1}+\omega_{2}\right)^{2}-\Delta^{2}\left(k_{1}+k_{2}\right)^{2}\right]\left[\left(k_{1}+k_{2}\right)^{2}+1\right]-\left(k_{1}+k_{2}\right)^{2}\right\} }\\
+\frac{\Delta^{2}k_{1}k_{2}\left[\left(k_{1}+k_{2}\right)^{2}+1\right]\left[\left(\omega_{1}+\omega_{2}\right)\tilde{A}_{01}\tilde{A}_{10}+\left(k_{1}+k_{2}\right)\left(\tilde{A}_{01}\tilde{B}_{10}+\tilde{A}_{10}\tilde{B}_{01}\right)\right]}{2\left\{ \left[\left(\omega_{1}+\omega_{2}\right)^{2}-\Delta^{2}\left(k_{1}+k_{2}\right)^{2}\right]\left[\left(k_{1}+k_{2}\right)^{2}+1\right]-\left(k_{1}+k_{2}\right)^{2}\right\} },
\end{multline}

\begin{multline}
\tilde{B}_{1,-1}=\frac{\left(\omega_{1}-\omega_{2}\right)k_{1}k_{2}\left[\left(k_{1}-k_{2}\right)^{2}+1\right]\tilde{B}_{01}\tilde{B}_{10}-\left(\omega_{1}-\omega_{2}\right)C_{01}C_{10}+k_{1}k_{2}\left(k_{1}-k_{2}\right)\left(\tilde{A}_{01}\tilde{B}_{10}+\tilde{A}_{10}\tilde{B}_{01}\right)}{2\left\{ \left[\left(\omega_{1}-\omega_{2}\right)^{2}-\Delta^{2}\left(k_{1}-k_{2}\right)^{2}\right]\left[\left(k_{1}-k_{2}\right)^{2}+1\right]-\left(k_{1}-k_{2}\right)^{2}\right\} }\\
+\frac{\Delta^{2}k_{1}k_{2}\left[\left(k_{1}-k_{2}\right)^{2}+1\right]\left[\left(\omega_{1}-\omega_{2}\right)\tilde{A}_{01}\tilde{A}_{10}+\left(k_{1}-k_{2}\right)\left(\tilde{A}_{01}\tilde{B}_{10}+\tilde{A}_{10}\tilde{B}_{01}\right)\right]}{2\left\{ \left[\left(\omega_{1}-\omega_{2}\right)^{2}-\Delta^{2}\left(k_{1}-k_{2}\right)^{2}\right]\left[\left(k_{1}-k_{2}\right)^{2}+1\right]-\left(k_{1}-k_{2}\right)^{2}\right\} },
\end{multline}

\begin{multline}
C_{11}=\frac{k_{1}k_{2}\left(k_{1}+k_{2}\right)^{2}\tilde{B}_{01}\tilde{B}_{10}-\left(\omega_{1}+\omega_{2}\right)^{2}C_{01}C_{10}+\left(\omega_{1}+\omega_{2}\right)k_{1}k_{2}\left(k_{1}+k_{2}\right)\left(\tilde{A}_{01}\tilde{B}_{10}+\tilde{A}_{10}\tilde{B}_{01}\right)}{2\left\{ \left[\left(\omega_{1}+\omega_{2}\right)^{2}-\Delta^{2}\left(k_{1}+k_{2}\right)^{2}\right]\left[\left(k_{1}+k_{2}\right)^{2}+1\right]-\left(k_{1}+k_{2}\right)^{2}\right\} }\\
+\frac{\Delta^{2}\left(k_{1}+k_{2}\right)^{2}\left(C_{01}C_{10}+k_{1}k_{2}\tilde{A}_{01}\tilde{A}_{10}\right)}{2\left\{ \left[\left(\omega_{1}+\omega_{2}\right)^{2}-\Delta^{2}\left(k_{1}+k_{2}\right)^{2}\right]\left[\left(k_{1}+k_{2}\right)^{2}+1\right]-\left(k_{1}+k_{2}\right)^{2}\right\} },
\end{multline}

\begin{multline}
C_{1,-1}=\frac{k_{1}k_{2}\left(k_{1}-k_{2}\right)^{2}\tilde{B}_{01}\tilde{B}_{10}-\left(\omega_{1}-\omega_{2}\right)^{2}C_{01}C_{10}+\left(\omega_{1}-\omega_{2}\right)k_{1}k_{2}\left(k_{1}-k_{2}\right)\left(\tilde{A}_{01}\tilde{B}_{10}+\tilde{A}_{10}\tilde{B}_{01}\right)}{2\left\{ \left[\left(\omega_{1}-\omega_{2}\right)^{2}-\Delta^{2}\left(k_{1}-k_{2}\right)^{2}\right]\left[\left(k_{1}-k_{2}\right)^{2}+1\right]-\left(k_{1}-k_{2}\right)^{2}\right\} }\\
+\frac{\Delta^{2}\left(k_{1}-k_{2}\right)^{2}\left(C_{01}C_{10}+k_{1}k_{2}\tilde{A}_{01}\tilde{A}_{10}\right)}{2\left\{ \left[\left(\omega_{1}-\omega_{2}\right)^{2}-\Delta^{2}\left(k_{1}-k_{2}\right)^{2}\right]\left[\left(k_{1}-k_{2}\right)^{2}+1\right]-\left(k_{1}-k_{2}\right)^{2}\right\} }.
\end{multline}

Finally, by using Eqs.~(\ref{eq:A_lin_1})\textendash (\ref{eq:B_lin_2}),
we can express all the second-order amplitudes through $C_{10}$ and
$C_{01}$:

\begin{equation}
\tilde{A}_{20}=\frac{k_{1}}{8}\frac{k_{1}^{2}\left(3\omega_{1}^{2}+\Delta^{2}k_{1}^{2}\right)\left(4k_{1}^{2}+1\right)-\left(\omega_{1}^{2}-\Delta^{2}k_{1}^{2}\right)^{2}}{\left(\omega_{1}^{2}-\Delta^{2}k_{1}^{2}\right)^{2}\left[\left(\omega_{1}^{2}-\Delta^{2}k_{1}^{2}\right)\left(4k_{1}^{2}+1\right)-k_{1}^{2}\right]}C_{10}^{2},\label{eq:A20_final}
\end{equation}

\begin{equation}
\tilde{A}_{02}=\frac{k_{2}}{8}\frac{k_{2}^{2}\left(3\omega_{2}^{2}+\Delta^{2}k_{2}^{2}\right)\left(4k_{2}^{2}+1\right)-\left(\omega_{2}^{2}-\Delta^{2}k_{2}^{2}\right)^{2}}{\left(\omega_{2}^{2}-\Delta^{2}k_{2}^{2}\right)^{2}\left[\left(\omega_{2}^{2}-\Delta^{2}k_{2}^{2}\right)\left(4k_{2}^{2}+1\right)-k_{2}^{2}\right]}C_{01}^{2},\label{eq:A02_final}
\end{equation}

\begin{equation}
\tilde{B}_{20}=\frac{\omega_{1}}{8}\frac{k_{1}^{2}\left(\omega_{1}^{2}+3\Delta^{2}k_{1}^{2}\right)\left(4k_{1}^{2}+1\right)+2k_{1}^{4}-\left(\omega_{1}^{2}-\Delta^{2}k_{1}^{2}\right)^{2}}{\left(\omega_{1}^{2}-\Delta^{2}k_{1}^{2}\right)^{2}\left[\left(\omega_{1}^{2}-\Delta^{2}k_{1}^{2}\right)\left(4k_{1}^{2}+1\right)-k_{1}^{2}\right]}C_{10}^{2},\label{eq:B20_final}
\end{equation}

\begin{equation}
\tilde{B}_{02}=\frac{\omega_{2}}{8}\frac{k_{2}^{2}\left(\omega_{2}^{2}+3\Delta^{2}k_{2}^{2}\right)\left(4k_{2}^{2}+1\right)+2k_{2}^{4}-\left(\omega_{2}^{2}-\Delta^{2}k_{2}^{2}\right)^{2}}{\left(\omega_{2}^{2}-\Delta^{2}k_{2}^{2}\right)^{2}\left[\left(\omega_{2}^{2}-\Delta^{2}k_{2}^{2}\right)\left(4k_{2}^{2}+1\right)-k_{2}^{2}\right]}C_{01}^{2},\label{eq:B02_final}
\end{equation}

\begin{equation}
C_{20}=\frac{k_{1}^{4}\left(3\omega_{1}^{2}+\Delta^{2}k_{1}^{2}\right)-\left(\omega_{1}^{2}-\Delta^{2}k_{1}^{2}\right)^{3}}{4\left(\omega_{1}^{2}-\Delta^{2}k_{1}^{2}\right)^{2}\left[\left(\omega_{1}^{2}-\Delta^{2}k_{1}^{2}\right)\left(4k_{1}^{2}+1\right)-k_{1}^{2}\right]}C_{10}^{2},\label{eq:C20_final}
\end{equation}

\begin{equation}
C_{02}=\frac{k_{2}^{4}\left(3\omega_{2}^{2}+\Delta^{2}k_{2}^{2}\right)-\left(\omega_{2}^{2}-\Delta^{2}k_{2}^{2}\right)^{3}}{4\left(\omega_{2}^{2}-\Delta^{2}k_{2}^{2}\right)^{2}\left[\left(\omega_{2}^{2}-\Delta^{2}k_{2}^{2}\right)\left(4k_{2}^{2}+1\right)-k_{2}^{2}\right]}C_{01}^{2},\label{eq:C02_final}
\end{equation}

\begin{multline}
\tilde{A}_{11}=\frac{k_{1}k_{2}\left[\left(\omega_{1}\omega_{2}+\Delta^{2}k_{1}k_{2}\right)\left(k_{1}+k_{2}\right)+\left(\omega_{1}+\omega_{2}\right)\left(k_{2}\omega_{1}+k_{1}\omega_{2}\right)\right]\left[\left(k_{1}+k_{2}\right)^{2}+1\right]}{2\left(\omega_{1}^{2}-\Delta^{2}k_{1}^{2}\right)\left(\omega_{2}^{2}-\Delta^{2}k_{2}^{2}\right)\left\{ \left[\left(\omega_{1}+\omega_{2}\right)^{2}-\Delta^{2}\left(k_{1}+k_{2}\right)^{2}\right]\left[\left(k_{1}+k_{2}\right)^{2}+1\right]-\left(k_{1}+k_{2}\right)^{2}\right\} }C_{01}C_{10}\\
-\frac{k_{1}+k_{2}}{2\left\{ \left[\left(\omega_{1}+\omega_{2}\right)^{2}-\Delta^{2}\left(k_{1}+k_{2}\right)^{2}\right]\left[\left(k_{1}+k_{2}\right)^{2}+1\right]-\left(k_{1}+k_{2}\right)^{2}\right\} }C_{01}C_{10},\label{eq:A11_final}
\end{multline}

\begin{multline}
\tilde{A}_{1,-1}=\frac{k_{1}k_{2}\left[\left(\omega_{1}\omega_{2}+\Delta^{2}k_{1}k_{2}\right)\left(k_{1}-k_{2}\right)+\left(\omega_{1}-\omega_{2}\right)\left(k_{2}\omega_{1}+k_{1}\omega_{2}\right)\right]\left[\left(k_{1}-k_{2}\right)^{2}+1\right]}{2\left(\omega_{1}^{2}-\Delta^{2}k_{1}^{2}\right)\left(\omega_{2}^{2}-\Delta^{2}k_{2}^{2}\right)\left\{ \left[\left(\omega_{1}-\omega_{2}\right)^{2}-\Delta^{2}\left(k_{1}-k_{2}\right)^{2}\right]\left[\left(k_{1}-k_{2}\right)^{2}+1\right]-\left(k_{1}-k_{2}\right)^{2}\right\} }C_{01}C_{10}\\
-\frac{k_{1}-k_{2}}{2\left\{ \left[\left(\omega_{1}-\omega_{2}\right)^{2}-\Delta^{2}\left(k_{1}-k_{2}\right)^{2}\right]\left[\left(k_{1}-k_{2}\right)^{2}+1\right]-\left(k_{1}-k_{2}\right)^{2}\right\} }C_{01}C_{10},\label{eq:A1-1_final}
\end{multline}

\begin{multline}
\tilde{B}_{11}=\frac{k_{1}k_{2}\left\{ \left[\left(\omega_{1}+\omega_{2}\right)\left(\omega_{1}\omega_{2}+\Delta^{2}k_{1}k_{2}\right)+\Delta^{2}\left(k_{1}+k_{2}\right)\left(k_{2}\omega_{1}+k_{1}\omega_{2}\right)\right]\left[\left(k_{1}+k_{2}\right)^{2}+1\right]+\left(k_{1}+k_{2}\right)\left(k_{2}\omega_{1}+k_{1}\omega_{2}\right)\right\} }{2\left(\omega_{1}^{2}-\Delta^{2}k_{1}^{2}\right)\left(\omega_{2}^{2}-\Delta^{2}k_{2}^{2}\right)\left\{ \left[\left(\omega_{1}+\omega_{2}\right)^{2}-\Delta^{2}\left(k_{1}+k_{2}\right)^{2}\right]\left[\left(k_{1}+k_{2}\right)^{2}+1\right]-\left(k_{1}+k_{2}\right)^{2}\right\} }\\
\times C_{01}C_{10}\\
-\frac{\omega_{1}+\omega_{2}}{2\left\{ \left[\left(\omega_{1}+\omega_{2}\right)^{2}-\Delta^{2}\left(k_{1}+k_{2}\right)^{2}\right]\left[\left(k_{1}+k_{2}\right)^{2}+1\right]-\left(k_{1}+k_{2}\right)^{2}\right\} }C_{01}C_{10},\label{eq:B11_final}
\end{multline}

\begin{multline}
\tilde{B}_{1,-1}=\frac{k_{1}k_{2}\left\{ \left[\left(\omega_{1}-\omega_{2}\right)\left(\omega_{1}\omega_{2}+\Delta^{2}k_{1}k_{2}\right)+\Delta^{2}\left(k_{1}-k_{2}\right)\left(k_{2}\omega_{1}+k_{1}\omega_{2}\right)\right]\left[\left(k_{1}-k_{2}\right)^{2}+1\right]+\left(k_{1}-k_{2}\right)\left(k_{2}\omega_{1}+k_{1}\omega_{2}\right)\right\} }{2\left(\omega_{1}^{2}-\Delta^{2}k_{1}^{2}\right)\left(\omega_{2}^{2}-\Delta^{2}k_{2}^{2}\right)\left\{ \left[\left(\omega_{1}-\omega_{2}\right)^{2}-\Delta^{2}\left(k_{1}-k_{2}\right)^{2}\right]\left[\left(k_{1}-k_{2}\right)^{2}+1\right]-\left(k_{1}-k_{2}\right)^{2}\right\} }\\
\times C_{01}C_{10}\\
-\frac{\omega_{1}-\omega_{2}}{2\left\{ \left[\left(\omega_{1}-\omega_{2}\right)^{2}-\Delta^{2}\left(k_{1}-k_{2}\right)^{2}\right]\left[\left(k_{1}-k_{2}\right)^{2}+1\right]-\left(k_{1}-k_{2}\right)^{2}\right\} }C_{01}C_{10},\label{eq:B1-1_final}
\end{multline}

\begin{multline}
C_{11}=\frac{k_{1}k_{2}\left(k_{1}+k_{2}\right)\left[\left(k_{1}+k_{2}\right)\left(\omega_{1}\omega_{2}+\Delta^{2}k_{1}k_{2}\right)+\left(\omega_{1}+\omega_{2}\right)\left(k_{2}\omega_{1}+k_{1}\omega_{2}\right)\right]}{2\left(\omega_{1}^{2}-\Delta^{2}k_{1}^{2}\right)\left(\omega_{2}^{2}-\Delta^{2}k_{2}^{2}\right)\left\{ \left[\left(\omega_{1}+\omega_{2}\right)^{2}-\Delta^{2}\left(k_{1}+k_{2}\right)^{2}\right]\left[\left(k_{1}+k_{2}\right)^{2}+1\right]-\left(k_{1}+k_{2}\right)^{2}\right\} }C_{01}C_{10}\\
-\frac{\left(\omega_{1}+\omega_{2}\right)^{2}-\Delta^{2}\left(k_{1}+k_{2}\right)^{2}}{2\left\{ \left[\left(\omega_{1}+\omega_{2}\right)^{2}-\Delta^{2}\left(k_{1}+k_{2}\right)^{2}\right]\left[\left(k_{1}+k_{2}\right)^{2}+1\right]-\left(k_{1}+k_{2}\right)^{2}\right\} }C_{01}C_{10},\label{eq:C11_final}
\end{multline}

\begin{multline}
C_{1,-1}=\frac{k_{1}k_{2}\left(k_{1}-k_{2}\right)\left[\left(k_{1}-k_{2}\right)\left(\omega_{1}\omega_{2}+\Delta^{2}k_{1}k_{2}\right)+\left(\omega_{1}-\omega_{2}\right)\left(k_{2}\omega_{1}+k_{1}\omega_{2}\right)\right]}{2\left(\omega_{1}^{2}-\Delta^{2}k_{1}^{2}\right)\left(\omega_{2}^{2}-\Delta^{2}k_{2}^{2}\right)\left\{ \left[\left(\omega_{1}-\omega_{2}\right)^{2}-\Delta^{2}\left(k_{1}-k_{2}\right)^{2}\right]\left[\left(k_{1}-k_{2}\right)^{2}+1\right]-\left(k_{1}-k_{2}\right)^{2}\right\} }C_{01}C_{10}\\
-\frac{\left(\omega_{1}-\omega_{2}\right)^{2}-\Delta^{2}\left(k_{1}-k_{2}\right)^{2}}{2\left\{ \left[\left(\omega_{1}-\omega_{2}\right)^{2}-\Delta^{2}\left(k_{1}-k_{2}\right)^{2}\right]\left[\left(k_{1}-k_{2}\right)^{2}+1\right]-\left(k_{1}-k_{2}\right)^{2}\right\} }C_{01}C_{10}.\label{eq:C1-1_final}
\end{multline}

Note that in the limiting case of cold ions $\left(\Delta=0\right)$
and with two identical counter propagating traveling waves (i.e.,
a standing wave with $k_{1}=-k,k_{2}=k$, $\omega_{1}=\omega_{2}=\omega$,
$\varepsilon_{1}=\varepsilon_{2}=\varepsilon$) the above amplitudes
coincide with the amplitudes derived in Ref.~\citep{Friedland2019},
as expected. In particular, the amplitudes $a_{1}$, $a_{2}$, $A$,
$b_{1}$, $b_{2}$, $c_{0}$, $c_{1}$, $c_{2}$, $C$ from Ref.~\citep{Friedland2019}
are related to the amplitudes in this paper as follows:

\begin{gather}
\tilde{A}_{10}=-\frac{a_{1}}{2},\tilde{A}_{01}=\frac{a_{1}}{2},\tilde{A}_{20}=-\frac{a_{2}}{2},\tilde{A}_{02}=\frac{a_{2}}{2},\tilde{A}_{1,-1}=-A,\\
\tilde{B}_{10}=-\frac{b_{1}}{2},\tilde{B}_{01}=-\frac{b_{1}}{2},\tilde{B}_{20}=-\frac{b_{2}}{2},\tilde{B}_{02}=-\frac{b_{2}}{2},\\
C_{00}=\frac{c_{0}}{2},C_{11}=\frac{c_{0}}{2},C_{10}=\frac{c_{1}}{2},C_{01}=\frac{c_{1}}{2},C_{1,-1}=C,C_{20}=\frac{c_{2}}{2},C_{02}=\frac{c_{2}}{2}.
\end{gather}

To derive weakly nonlinear equations we also need to calculate the
variations of $\bar{L}$ with respect to the first-order amplitudes:

\begin{equation}
\frac{\partial\bar{L}}{\partial C_{10}}=-\frac{1}{2}k_{1}\tilde{A}_{10}+\frac{1}{2}\left(1+k_{1}^{2}\right)C_{10}+\frac{1}{2}C_{10}\left(\frac{1}{8}C_{10}^{2}+\frac{1}{4}C_{01}^{2}+C_{00}+\frac{1}{2}C_{20}\right)+\frac{1}{4}C_{01}\left(C_{1,-1}+C_{11}\right)+\frac{\varepsilon_{1}}{2}\cos\left(\Phi_{1}\right)=0,\label{eq:C10}
\end{equation}

\begin{equation}
\frac{\partial\bar{L}}{\partial C_{01}}=-\frac{1}{2}k_{2}\tilde{A}_{01}+\frac{1}{2}\left(1+k_{2}^{2}\right)C_{01}+\frac{1}{2}C_{01}\left(\frac{1}{8}C_{01}^{2}+\frac{1}{4}C_{10}^{2}+C_{00}+\frac{1}{2}C_{02}\right)+\frac{1}{4}C_{10}\left(C_{1,-1}+C_{11}\right)+\frac{\varepsilon_{2}}{2}\cos\left(\Phi_{2}\right)=0,\label{eq:C01}
\end{equation}

\begin{multline}
\frac{\partial\bar{L}}{\partial\tilde{B}_{10}}=-\frac{1}{2}k_{1}^{2}\left(1+k_{1}\tilde{A}_{20}\right)\tilde{B}_{10}-\frac{1}{4}k_{1}k_{2}\left[\left(k_{1}-k_{2}\right)\tilde{A}_{1,-1}+\left(k_{1}+k_{2}\right)\tilde{A}_{11}\right]\tilde{B}_{01}\\
+\frac{1}{2}k_{1}\left(\omega_{1}-k_{1}^{2}\tilde{B}_{20}\right)\tilde{A}_{10}-\frac{1}{4}k_{1}k_{2}\left[\left(k_{1}-k_{2}\right)\tilde{B}_{1,-1}+\left(k_{1}+k_{2}\right)\tilde{B}_{11}\right]\tilde{A}_{01}=0,\label{eq:B10}
\end{multline}

\begin{multline}
\frac{\partial\bar{L}}{\partial\tilde{B}_{01}}=-\frac{1}{2}k_{2}^{2}\left(1+k_{2}\tilde{A}_{02}\right)\tilde{B}_{01}-\frac{1}{4}k_{1}k_{2}\left[\left(k_{1}-k_{2}\right)\tilde{A}_{1,-1}+\left(k_{1}+k_{2}\right)\tilde{A}_{11}\right]\tilde{B}_{10}\\
+\frac{1}{2}k_{2}\left(\omega_{2}-k_{2}^{2}\tilde{B}_{02}\right)\tilde{A}_{01}-\frac{1}{4}k_{1}k_{2}\left[\left(k_{1}-k_{2}\right)\tilde{B}_{1,-1}+\left(k_{1}+k_{2}\right)\tilde{B}_{11}\right]\tilde{A}_{10}=0,\label{eq:B01}
\end{multline}

\begin{multline}
\frac{\partial\bar{L}}{\partial\tilde{A}_{10}}=-\frac{1}{2}k_{1}C_{10}+\frac{1}{2}k_{1}\left(\omega_{1}-k_{1}^{2}\tilde{B}_{20}\right)\tilde{B}_{10}-\frac{1}{4}k_{1}k_{2}\left[\left(k_{1}-k_{2}\right)\tilde{B}_{1,-1}+\left(k_{1}+k_{2}\right)\tilde{B}_{11}\right]\tilde{B}_{01}\\
-\frac{\Delta^{2}}{2}k_{1}^{2}\left(1+k_{1}\tilde{A}_{20}\right)\tilde{A}_{10}-\frac{\Delta^{2}}{4}k_{1}k_{2}\left[\left(k_{1}-k_{2}\right)\tilde{A}_{1,-1}+\left(k_{1}+k_{2}\right)\tilde{A}_{11}\right]\tilde{A}_{01}=0,\label{eq:A10}
\end{multline}

\begin{multline}
\frac{\partial\bar{L}}{\partial\tilde{A}_{01}}=-\frac{1}{2}k_{2}C_{01}+\frac{1}{2}k_{2}\left(\omega_{2}-k_{2}^{2}\tilde{B}_{02}\right)\tilde{B}_{01}-\frac{1}{4}k_{1}k_{2}\left[\left(k_{1}-k_{2}\right)\tilde{B}_{1,-1}+\left(k_{1}+k_{2}\right)\tilde{B}_{11}\right]\tilde{B}_{10}\\
-\frac{\Delta^{2}}{2}k_{2}^{2}\left(1+k_{2}\tilde{A}_{02}\right)\tilde{A}_{01}-\frac{\Delta^{2}}{4}k_{1}k_{2}\left[\left(k_{1}-k_{2}\right)\tilde{A}_{1,-1}+\left(k_{1}+k_{2}\right)\tilde{A}_{11}\right]\tilde{A}_{10}=0.\label{eq:A01}
\end{multline}

\section{Functions $P\left(k_{1},\omega_{1}\right)$ and $Q\left(k_{1},\omega_{1};k_{2},\omega_{2}\right)$} 
\label{Appendix_C}

Function $P\left(k_{1},\omega_{1}\right)$ is defined through

\begin{equation}
C_{10}^{2}P\left(k_{1},\omega_{1}\right)=-\frac{1}{8}C_{10}^{2}+\frac{1}{2}C_{20}-\frac{2\omega_{1}k_{1}^{4}}{\left(\omega_{1}^{2}-\Delta^{2}k_{1}^{2}\right)^{2}}\tilde{B}_{20}-\frac{k_{1}^{3}\left(\omega_{1}^{2}+\Delta^{2}k_{1}^{2}\right)}{\left(\omega_{1}^{2}-\Delta^{2}k_{1}^{2}\right)^{2}}\tilde{A}_{20},
\end{equation}

\noindent or, equivalently due to symmetry, through

\begin{equation}
C_{01}^{2}P\left(k_{2},\omega_{2}\right)=-\frac{1}{8}C_{01}^{2}+\frac{1}{2}C_{02}-\frac{2\omega_{2}k_{2}^{4}}{\left(\omega_{2}^{2}-\Delta^{2}k_{2}^{2}\right)^{2}}\tilde{B}_{02}-\frac{k_{2}^{3}\left(\omega_{2}^{2}+\Delta^{2}k_{2}^{2}\right)}{\left(\omega_{2}^{2}-\Delta^{2}k_{2}^{2}\right)^{2}}\tilde{A}_{02},
\end{equation}

\noindent while $Q\left(k_{1},\omega_{1};k_{2},\omega_{2}\right)$
is defined through

\begin{multline}
C_{01}C_{10}Q\left(k_{1},\omega_{1};k_{2},\omega_{2}\right)=\frac{1}{2}\Biggl\{ C_{1,-1}+C_{11}-\frac{k_{1}k_{2}\left(\omega_{1}\omega_{2}+\Delta^{2}k_{1}k_{2}\right)}{\left(\omega_{1}^{2}-\Delta^{2}k_{1}^{2}\right)\left(\omega_{2}^{2}-\Delta^{2}k_{2}^{2}\right)}\left[\left(k_{1}-k_{2}\right)\tilde{A}_{1,-1}+\left(k_{1}+k_{2}\right)\tilde{A}_{11}\right]\\
-\frac{k_{1}k_{2}\left(k_{1}\omega_{2}+k_{2}\omega_{1}\right)}{\left(\omega_{1}^{2}-\Delta^{2}k_{1}^{2}\right)\left(\omega_{2}^{2}-\Delta^{2}k_{2}^{2}\right)}\left[\left(k_{1}-k_{2}\right)\tilde{B}_{1,-1}+\left(k_{1}+k_{2}\right)\tilde{B}_{11}\right]\Biggr\}.
\end{multline}

Here, the amplitudes $\tilde{A}_{20}$, $\tilde{A}_{02}$, $\tilde{B}_{20}$,
$\tilde{B}_{02}$, $C_{20}$, $C_{02}$, $\tilde{A}_{11}$, $\tilde{A}_{1,-1}$,
$\tilde{B}_{11}$, $\tilde{B}_{1,-1}$, $C_{11}$, $C_{1,-1}$ should
be expressed through $C_{10}$, $C_{01}$, $k_{1}$, $k_{2}$, $\omega_{1}$,
$\omega_{2}$, $\Delta$ using Eqs.~(\ref{eq:A20_final})\textendash (\ref{eq:C1-1_final}),
so that $P\left(k_{1},\omega_{1}\right)$, $Q\left(k_{1},\omega_{1};k_{2},\omega_{2}\right)$
are functions of $k_{1}$, $k_{2}$, $\omega_{1}$, $\omega_{2}$,
$\Delta$ only.

\end{widetext}

\end{document}